%% file: Paper.tex
\documentclass[preprint, 5p, twocolumn]{elsarticle}

\usepackage{amsmath,amssymb,amsfonts}
\usepackage{algorithmic}
\usepackage{graphicx}
\usepackage{textcomp}
\usepackage{xcolor}
\usepackage{microtype}
\usepackage{todonotes}
\usepackage{framed}
\usepackage{pgfplots}
\usepackage{tikz}
\usetikzlibrary{shapes,arrows}
\usepackage{pgf-pie}
\usepackage{soul}
\usepackage{array}
\usepackage{url}
\usepackage{listings, multicol}
\usepackage[normalem]{ulem}
\usepackage[bookmarks=false]{hyperref}
\usepackage{adjustbox}
\usepackage{tabu}
\usepackage{paralist}
\usepackage{subcaption}
\usepackage{verbatim}
\usepackage{float}
\usepackage{multirow} 
\usepackage{lscape}
\usepackage{booktabs}
\usepackage{xspace}
\usepackage{enumitem}
\usepackage{color}
\usepackage{pifont}
\let\url\nolinkurl
\graphicspath{ {images/} }

\newcommand{\JS}{Java\-Script\xspace}

\newcommand{\eJS}{\emph{evaluateJavascript}\xspace}

\newcommand{\cmark}{\ding{51}}
\newcommand{\xmark}{\ding{55}}

\journal{Journal of Systems and Software}

\begin{document}

\begin{frontmatter}

\title{A Large Scale Analysis of Android -- Web Hybridization\tnoteref{t1}}
\tnotetext[t1]{ This work was partially supported by the German Federal Ministry of Education and Research (BMBF) through the project SmartPriv (16KIS0760) and the German Research Foundation (DFG) via the collaborative research center ``Methods and Tools for Understanding and Controlling Privacy'' (SFB 1223), project B02.\newline
}

\author[1]{Abhishek Tiwari\fnref{nus}}
\ead{tiwari@comp.nus.edu.sg}
\address[1]{National University of Singapore, Singapore City, Singapore}

\author[2]{Jyoti\corref{cor} Prakash}
\ead{jyoti@uni-potsdam.de}
\cortext[cor]{Corresponding Author}
\author[2]{Sascha Gro{\ss}}
\ead{gross@uni-potsdam.de}
\author[2]{Christian Hammer}
\ead{hammer@cs.uni-potsdam.de}
\address[2]{University of Potsdam, Potsdam, Germany}
\fntext[nus]{While performing this work the author was at the University of Potsdam}


\definecolor{dkgreen}{rgb}{0,0.6,0}
\definecolor{gray}{rgb}{0.5,0.5,0.5}
\definecolor{mauve}{rgb}{0.58,0,0.82}

\lstdefinelanguage{Java}{
  keywords={typeof, new, true, false, catch, function, return, null, catch, switch, var, if, in, while, do, else, case, break,class, export, boolean, throw, implements, import, this},
  morekeywords = {public, private, protected},
  keywordstyle=\color{blue},
  commentstyle=\color{green!40!black},
  stringstyle=\color{violet!90},
  basicstyle={\footnotesize\sffamily},
  comment=[l]{//},
  morecomment=[s]{/*}{*/},
  morestring=[b]',
  morestring=[b]",
  frame=tb,
  aboveskip=3mm,
  belowskip=3mm,
  showstringspaces=false,
  columns=flexible,
  numbers=left,
  numberstyle=\tiny\color{black},
  breaklines=true,
  breakatwhitespace=true,
  tabsize=2,
  numbersep=3pt
}


\lstdefinelanguage{JavaScript}{
  keywords={typeof, new, true, false, catch, function, return, null, catch, switch, var, if, in, while, do, else, case, break,class, export, boolean, throw, implements, import, this},
  keywordstyle=\color{blue},classoffset=1,
  morekeywords = {SynchJS, HTMLOUT, Appnext, document, Sponsorpay},
  commentstyle=\color{green!40!black},
  stringstyle=\color{violet!90},
  basicstyle={\footnotesize\sffamily},
  identifierstyle=\color{black},
  sensitive=false,
  comment=[l]{//},
  morecomment=[s]{/*}{*/},
  morestring=[b]',
  morestring=[b]",
  frame=tb,
  columns=flexible,
  numbers=left,
  numberstyle=\tiny\color{black},
  escapeinside={(*}{*)},
  breaklines=true,
  breakatwhitespace=true,
  tabsize=2,
  numbersep=3pt
}

\begin{abstract}
\input{sections/abstract.tex}
\end{abstract}


\end{frontmatter}

\section{Introduction}
\input{sections/introduction.tex}

\section{Background}
\input{sections/background.tex}

\section{Methodology}
\input{sections/methodology.tex}

\section{Dataset Selection}
\input{sections/dataset.tex}

\section{Evaluation --- IFC and URL Analysis}
\input{sections/evaluation.tex}

\section{Evaluation --- \JS in Benign Apps}
\input{sections/evaluation-JS.tex}

\section{Evaluation --- \JS usage in frequently used apps}
\input{sections/evaluation-js-cateogry}


\section{Evaluation --- \JS usage in Malware}\label{eval-malware}
\input{sections/evaluation-malware.tex}

\section{Threats to Validity}
\input{sections/discussion.tex}

\section{Related Work}
\input{sections/related-work.tex}
	
\section{Conclusion}
\input{sections/conclusion.tex}


\bibliographystyle{elsarticle-num}
\bibliography{Paper}



\end{document}

%% file: sections/abstract.tex
%
Many Android applications embed webpages via Web\-View components and execute \JS code within Android. Hybrid applications leverage dedicated APIs to load a resource and render it in a WebView. Furthermore, Android objects can be shared with the \JS world. However, bridging the interfaces of the Android and \JS world might also incur severe security threats: Potentially untrusted webpages and their \JS might interfere with the Android environment and its access to native features.

No general analysis is currently available to assess the implications of such hybrid apps bridging the two worlds. To understand the semantics and effects of hybrid apps, we perform a large-scale study on the usage of the hybridization APIs in the wild. We analyze and categorize the parameters to hybridization APIs for 7,500 randomly selected and the 196 most popular applications from the Google Playstore as well as 1000 malware samples. 
Our results advance the general understanding of hybrid applications, as well as implications for potential program analyses, and the current security situation: We discovered thousands of flows of sensitive data from Android to \JS, the vast majority of which could flow to potentially untrustworthy code. Our analysis identified numerous web pages embedding vulnerabilities, which we exemplarily exploited. Additionally, we discovered a multitude of applications in which potentially untrusted \JS code may interfere with (trusted) Android objects, both in benign and malign applications.

%% file: sections/introduction.tex
The usage of mobile devices is rapidly growing with Android being the most prevalent mobile operating system (global market share of 72.23\% as of November~2018~\cite{stats3}). Various reports~\cite{stats},\cite{stats1} reveal that mobile application (app) usage is growing by 6\% year-over-year and users are preferring mobile apps over desktop apps.

Considering these statistics, industry prioritizes mobile app development~\cite{stats4}. However, apps need to be developed for various platforms, such as Android and iOS, resulting in increased production time and cost. Traditional approaches require creation of a native application for each platform or of a universal web app. The former approach incurs redundant programming efforts, whereas, the latter suffers from the inability to access platform-specific information. 

Hybrid mobile apps combine native components with web components into a single mobile application. Intuitively, hybrid apps are native applications combined with web technologies such as HTML, JavaScript and CSS. On Android, a \emph{WebView}~\cite{webview} component, a chromeless browser~\cite{chromium} capable of displaying webpages, embeds the web applications into a view of the Android app. Ionic's developer survey~\cite{stats2} shows the increasing prevalence of hybrid applications. In previous years (2015-17), app development with native tools decreased significantly (by almost $7\!\times$), whereas the number of hybrid apps was growing in share of overall app development.

Due to the fact that hybrid apps combine native and web technologies in a single app, the attack surface for malicious activities increases significantly, as potentially untrusted code loaded at runtime can interfere with the trusted Android environment. In our study with 7,500 random applications from the Google Play Store, we found that 68\% of these apps use at least one instance of WebView and 87.9\% of these install an active communication channel between Android and JavaScript. This includes leaking various pieces of sensitive information, such as the user's location. To assess the impact on user privacy a standalone analysis of the Android or JavaScript side is thus clearly insufficient. However, very limited work towards linking the assessment of both worlds can be found in the literature. Lee et al.~\cite{lee2016hybridroid} provide a framework for hybrid communication's type error discovery and taint analysis of information flows between Android and JavaScript.  However, their framework  focuses on discovering type errors.  They evaluate their taint analysis on 48 apps only without providing insights into the nature of information flows from Android to \JS and vice versa. Besides, they overlook the (mis)usage of URLs and \JS inside hybridization APIs. Other related works only consider a very specific vulnerability arising from hybridization~\cite{codeinjection,cordova,rizzo2017babelview,hidhaya2015supplementary,mandal2018vulnerability,Fratantonio2016LogicBomb,yang2018study}

Hybrid communication leverages APIs like the \emph{loadUrl} or \emph{evaluateJavascript} methods, which, from inside an Android application, can either load a webpage into the WebView or execute JavaScript code directly. To improve comprehension of hybrid communication we performed a study, LUDroid, using a framework supporting our semi-automatic analysis of hybrid communication on Android. We extract the information flows from Android to hybridization APIs and thus to the JavaScript engine, to be executed in the displayed web page (if any), and categorize these flows into benign and transmitting sensitive data. We discovered 6,375 sensitive flows from Android to JavaScript.

The major parameter passed to \emph{loadUrl} is the URL to be loaded. We analyzed the syntax and semantics of each URL and provide a detailed categorization. As a byproduct, we found several vulnerabilities concerning the usage of these URLs. We successfully exploited some of these to demonstrate the threats. 

Alternatively, \emph{loadUrl} and \emph{evaluateJavascript} accept raw JavaScript code as a parameter. We encountered code that loads additional JavaScript libraries into WebView. Unexpectedly, we also discovered 653 applications (with potentially untrusted JavaScript code) transmitting data back to the Android bridge object. Therefore these apps implement two-way communication that may jeopardize the integrity of the Android environment, particularly as most external JavaScript is loaded without \emph{https} and is thus prone to man-in-the-middle attacks. We found that many apps save the runtime state of WebView to Android before destroying that component. These apps perform various privacy leaking functions. These functions include: (1) fingerprinting the device for advertisement or monetization purposes and (2) obfuscating code to preclude analysis of its semantics.
We discuss the impact of our findings on potential program analyses that are to automatically identify issues of hybrid apps while taking hybrid communication concisely into account.

Additional studies with the most popular 196 apps from the Google Play Store and 1000 malware samples from the AMD dataset~\cite{amdMalwares} show differing properties than those by the general benign apps. The malware ignores \eJS but -- even though many are repackaged benign apps -- leverage many SDKs, potentially for financial gain. Clickjacking attacks have a similar motivation. Other scenarios involve information theft or injection attacks. In contrast the most popular apps display a peculiar usage of \eJS, which differs from other apps, e.g., to control the page navigation or to inject secrets into web forms (however, these secrets are stored in plain text in the apk.) We also found that security-critical popular apps like banking do not use \emph{loadURL}. Many popular apps leverage SDKs from social networks or mobile payment.

Technically we provide the following contribution:
\begin{itemize}
\item \emph{Information Flow Analysis} We thoroughly investigate around 8,700 real-world Android apps, both benign and malign. We provide statistics on the information flows between Android and JavaScript and identify leakage of sensitive information to the WebView.
\item \emph{URL Analysis} We perform an extensive analysis for the URLs used with \emph{loadURL}, extracting various features. 
As a byproduct we identify applications that are vulnerable due to unencrypted transport protocols and exemplify the simplicity of a phishing attack.
\item \emph{JavaScript analysis} We inspect the JavaScript code passed to the hybridization APIs and identify a much smaller set distinct code snippets, most of which originating from third-part libraries. We compare the behaviour of regular benign apps to the most popular apps and malware and find several significant differences and security flaws. We extract relevant features and highlight their implications on program analysis of hybrid apps.
\end{itemize}

This paper is an extension of a conference paper~\cite{tiwari19SCAM}. This extended version adds the following main contributions: 
\begin{compactitem}
\item Support for evaluateJavaScript (section~\ref{section:top-100-js})
\item Extension of the evaluation dataset to include an
\begin{compactitem}
\item Analysis of the top 196 apps from the top 11 categories in Google Play Store (section~\ref{top100-loadURL})
\item Analysis of 1000 samples from 71 families of malware from the Argus AMD malware dataset (section~\ref{eval-malware})
\end{compactitem}
\item Comparison of the data usage between malware and benign apps (section~\ref{eval-malware})
\item Brief explanation of the methodology (section~\ref{methodology-ifc}).
\end{compactitem}

To summarize, our work advances the \emph{state-of-the-art} in understanding the usage in Android-Web hybridization and elucidates their relevant implications on program analysis, particularly for security and/or privacy scenarios. However, the aim of this work is not to discover specific vulnerabilities in Android-Web hybridization, but to provide a set of use cases to validate future implementations of program analyses for hybrid apps.

%% file: sections/background.tex

\subsection{Hybrid applications} A general disadvantage of native applications is that they are bound to a specific platform. For instance an Android application is bound to the Android platform and cannot easily be transformed into an iOS application. A developer wanting to support multiple platforms needs to implement a native application for each of these platforms separately, multiplying the implementation effort. Alternatively, web applications execute in an arbitrary web browser and are therefore platform independent. However, they are restricted by a browser sandbox with very limited access to the native APIs of the mobile device. \emph{Hybrid applications} have been proposed as a remedy, as they take full advantage of both approaches. They make extensive use of web requests, e.g., to display user interfaces. While having access to all native API methods granted by the Android permission system, development effort is reduced, as the user interface and its controllers can be retrieved via web requests and therefore do not need to be re-implemented.

\subsection{WebView, loadURL, and evaluateJavaScript} Hybrid applications on Android leverage \emph{WebView}s. WebViews are user interface components that display webpages (without any browser bars), and thus provide a means to implement user interfaces as web pages rather than natively. The WebView class provides a \emph{loadURL} method, which loads a webpage or executes raw \JS. This method comes in two variants: \emph{loadUrl(String url)} and \emph{loadUrl(String url, Map$<$String,String$>$ additionalHttpHeaders)}. Additional to the URL provided as argument to the first method, the second variant accepts additional headers for the HTTP request. Similar to a browser's location bar, one of the following parameters can be passed to \emph{loadUrl}: (1) a remote URL
leveraging protocols such as HTTP(S), (2) a local URL specified with protocol \emph{file}, or
(3) \JS code via the pseudo-protocol \emph{javascript:}. 
\emph{WebView} leverages the appropriate renderer for each URL type automatically. Finally, a dedicated API \emph{evaluateJavascript} executes \JS code directly.

\section{Motivating example}
\input{sections/background-motivational-example}

%% file: sections/background-motivational-example.tex

In this section we will describe a simple hybrid Android example program (Listing~\ref{listing:MainActivity1} and~\ref{listing:Leaker}) together with the communication between Android and the WebView component. We will then motivate the rationale behind our large-scale study to understand various factors concerning the usage of \emph{WebView}s in realistic apps. 

In Listing~\ref{listing:MainActivity1}, a \emph{WebView} object \emph{myWebView} is retrieved from the Activity's  UI via an identifier (line~\ref{lst:line1_2}). Execution of \JS in a \emph{WebView} object is disabled by default but can be enabled by overriding its default settings (line~\ref{lst:line1_5}). A Java interface object can be shared with the WebView to be accessible via \JS. Via this object the capabilities of the Android world can be bridged to the Web component. This \emph{bridge communication} allows \JS to, e.g., access various sensors' data that are usually only accessible from Android. In our example, an object of the class \emph{Leaker} (see Listing~\ref{listing:Leaker}) is shared (Listing~\ref{listing:MainActivity1}, line~\ref{lst:line1_7}) with the \emph{WebView} object \emph{myWebView}, such that every webpage loaded into \emph{myWebView} can access this object via the global variable \emph{``Android"} (i.e.~\emph{``Android"} becomes a persistent property of the DOM's global object). Finally, the method \emph{loadUrl} can be used in two ways: (1) to invoke \JS code directly (prepending a \emph{javascript:} pseudo-protocol to the passed code) from Android, and (2) to load a custom URL (line~\ref{lst:line1_10}) (which may execute \JS code specified or loaded in the web page).
\begin{lstlisting}[ language=Java, belowskip=-0.8 \baselineskip, caption=MainActivity.java, label={listing:MainActivity1}, escapeinside={@}{@}, float=*]
@\label{lst:line1_0}@ protected void onCreate(Bundle savesInstanceState) {
@\label{lst:line1_2}@	WebView myWebView = (WebView) findViewById(R.id.webview);
@\label{lst:line1_3}@	WebSettings webSettings = myWebView.getSettings();
//enable JavaScript on WebView
@\label{lst:line1_5}@	webSettings.setJavaScriptEnabled(true); 
@\label{lst:line1_6}@ // add interface object of type Leaker to the WebView's DOM as a property named "Android" of the global object 
  Leaker obj = new Leaker(this);
@\label{lst:line1_7}@	myWebView.addJavascriptInterface(obj, "Android");
@\label{lst:line1_8}@  //case 1: invoke JavaScript from Android
@\label{lst:line1_9}@	 myWebView.loadUrl("javascript:"+" print(Android.showToast('Hello World'))");
@\label{lst:line1_10}@  //case 2: load a webpage (potentially executing JavaScript), the object "Android" persists as property of the DOM's global object 
@\label{lst:line1_11}@  myWebView.loadUrl( "http://www.dummypage.com");
}
\end{lstlisting}
\begin{lstlisting}[language=Java, belowskip=-0.8 \baselineskip, caption=Leaker.java, label={listing:Leaker}, float, escapeinside={*}{*}, float=*, breaklines=true]
*\label{lst:line2_1}*//Add a JavascriptInterface annotation before the method you want to bridge
*\label{lst:line2_2}*@JavascriptInterface
*\label{lst:line2_3}*public String showToast(String toast) {
*\label{lst:line2_4}*	TelephonyManager tManager = (TelephonyManager) mContext.getSystemService(Context. TELEPHONY_SERVICE);
*\label{lst:line2_5}*	String uid = tManager.getDeviceId(); // get the device ID
*%\label{lst:line2_6}	 //smsManager.sendTextMessage("1234567890", null, uid, null, null); //sink 
**\label{lst:line2_7}* return uid;
}
\end{lstlisting}

Previous work~\cite{lee2016hybridroid} take a first step into analyzing the data flows from Android to \JS but  is only partially sound, and concentrates on potential type errors when passing data between the two worlds. 
To improve the understanding of which data flows are to be considered when analyzing an app consisting of a combination of Android and \JS code a thorough understanding of the methods \emph{addJavascriptInterface}, \emph{loadUrl}, and \emph{evaluateJavascript} is required. In particular, we are interested in the uses and potential abuses of this interface in the wild and their implications on the design of a program analysis for hybrid apps. 

Consider line~\ref{lst:line1_9} in Listing~\ref{listing:MainActivity1}, which reveals that the \emph{loadUrl} method is invoking the \emph{showToast} method defined in the \emph{Leaker} class (Listing~\ref{listing:Leaker}, line~\ref{lst:line2_3}). This Java method retrieves the Android device's unique ID and returns it to the \JS code. Similar \JS code could also be invoked in the loaded webpage (Listing~\ref{listing:MainActivity1}, line~\ref{lst:line1_11}) where it might be leaked to some untrusted web server together with more information the user enters into the web page. Note that state-of-the-art information flow analyses for Android cannot report this to be an illicit information flow, as they have no information whether the WebView's code actually leaks the shared data (or do not even consider \emph{loadURL} a potential information sink~\cite{susi}). To further investigate this scenario, access to the executed \JS code is required. However, static analysis of \JS code is challenging due to its highly dynamic nature~\cite{JSchallenges}, and as it additionally requires a careful inspection of the various aspects of the \emph{WebView} class and its bridge mechanism. Therefore, it becomes critical for program analyses to fully understand the behavior of \emph{loadUrl}, \emph{evaluateJavascript} and \emph{addJavascriptInterface}. The aim of our work is to provide this information by performing a large-scale study of real-world apps.

%% file: sections/methodology.tex
\input{sections/methodology-overview.tex}

\subsection{IFCAnalyzer}\label{methodology-ifc}
\input{sections/methodology-JSIntAnalzyer.tex}

\subsection{URLAnalyzer}
\input{sections/methodology-urlanalyzer.tex}

\subsection{JSAnalyzer}\label{methodology-js}
\input{sections/methodology-jsanalyzer.tex}

%% file: sections/methodology-overview.tex
\begin{figure}[t!]
\begin{adjustbox}{width=\linewidth}
	\small
\includegraphics[width=\textwidth]{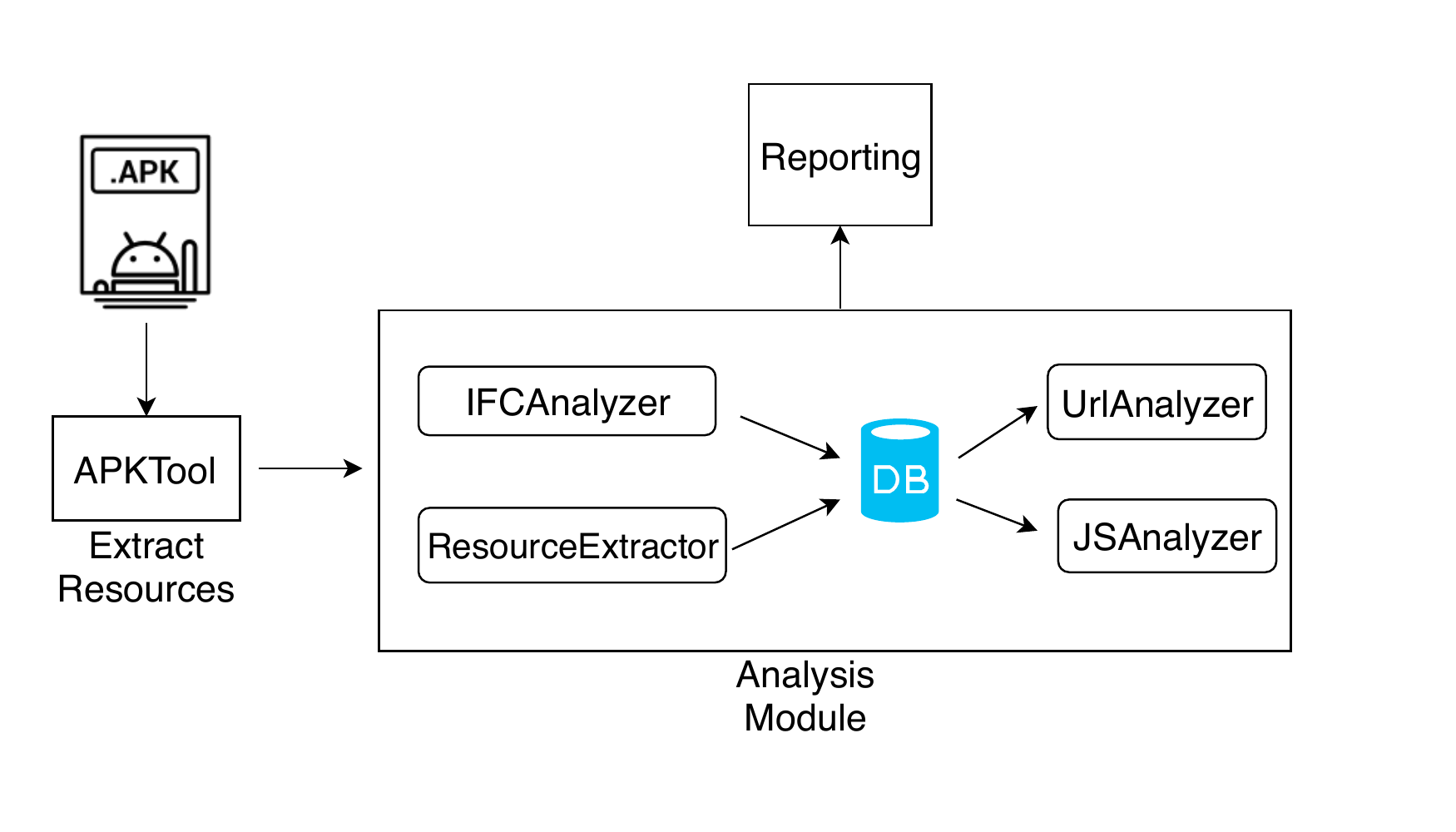}
\end{adjustbox}
\caption{The workflow of LUDroid}
\label{fig:workflow}
\end{figure}
We develop the toolchain \emph{LuDroid} to facilitate the semi-automated analysis for discovering the usage of \emph{loadUrl} and \eJS. Figure~\ref{fig:workflow} presents the workflow of LUDroid's analysis framework, consisting of the following modules: \emph{IFCAnalyzer}, \emph{ResourceExtractor}, \emph{UrlAnalyzer}, and \emph{JSAnalyzer}. LUDroid decompiles an APK\footnote{An APK is the binary format of an Android application.} using APKTool~\cite{apktool}. The decompiled output contains the app's resources as well as source code in the Smali~\cite{smali} format
. The \emph{IFCAnalyzer} module computes the set of statements that influence the method \emph{addJavascriptInterface} (i.e.~the \emph{backward slice}~\cite{weiser}). The backward slice is used to analyze which information flows from the Android to the \JS side. The \emph{ResourceExtractor} module extracts the strings that are passed as parameters to the two variants of the method \emph{loadUrl} and the method \emph{evaluateJavascript}. This data is stored in a database that is passed as input to the modules \emph{UrlAnalyzer} and \emph{JSAnalyzer}. The \emph{UrlAnalyzer} module analyzes the URLs provided as string argument to the two variants of the \emph{loadUrl} method. It validates the URLs and extracts various features, such as the used protocol, which facilitates the analysis of URLs in hybrid communication. Similarly, the \emph{JSAnalyzer} module analyzes the \JS code that is passed to the \emph{loadUrl} and \emph{evaluateJavascript} methods. In the followings we discuss each module in detail.

%% file: sections/methodology-JSIntAnalzyer.tex
The aim of this module is to understand the nature of the information flowing from Android to \JS. In particular we plan to answer the following research questions:
\begin{compactitem}
	\item\textbf{RQ1.1:} \textit{How pervasive is information flow from Android to \JS?}
	\item \textbf{RQ1.2:} \textit{Do these information flows include sensitive information?}
\end{compactitem}
We consider a piece of information to be sensitive if leaking it will violate its owner's privacy. Previous work has identified APIs that return potentially sensitive information~\cite{susi}, and we conservatively consider data sensitive if it originates from any of these information sources .

Figure~\ref{fig:jsworkflow} describes the workflow of the \emph{IFCAnalyzer} module. For every occurrence of the method \emph{addJavascriptInterface} we compute its backward slice to identify the corresponding \emph{WebView} initialization. The \emph{addJavascriptInterface} method injects an Android object into the target \emph{WebView}. It takes two parameters, a Java object and the name used to expose this object in this \emph{WebView}'s \JS engine. If \emph{\JS} is enabled in this \emph{WebView}, the loaded web pages can invoke the methods exposed by the injected Java object (cf. Listing~\ref{listing:Leaker}). As we are interested in the exposed functionality of this Java object, we extract its class and exposed methods: Not all methods of the Java object are bridged. Only methods annotated with \emph{@JavascriptInterface} are made available to \JS. We then identify (potential) sensitive information flows originating from these methods: \JS could invoke these methods to leak the returned sensitive information. Finally, we store the analysis results into a database.  

The backward slices leveraged by \emph{IFCAnalyzer} at the point of this writing are transitively back-tracing explicit information flows (i.e. definition-use chains) for the variables in question. For example, in Listing~\ref{listing:MainActivity1} our analysis computes a backward slice for the parameter \emph{obj} passed to \emph{addJavascriptInterface} and determines it to be of the class \emph{Leaker}. Further, a backward slice for its target object, \emph{myWebView}, returns the object defined in line~\ref{lst:line1_2}. Based on information computed via backward slicing, \emph{IFCAnalyzer} determines the bridge object, extracts the details of the annotated methods, and stores the results in a database.

Note that information could also be transmitted by implicit information flows (i.e. without def-use chains), however, implicit flows only allow to transmit information bit by bit, e.g.~in a loop. As we currently do not support string manipulation via character arrays (cf. section~\ref{ResourceExtractor}), it does not make sense to consider implicit flows when computing backward slices. However, we consider these topics important for a proper analysis that brides the \JS and Android worlds in order to preclude certain obfuscations of information flows.

\tikzstyle{decision} = [diamond, draw, fill=blue!20, 
    text width=5em, text badly centered, node distance=3cm, inner sep=1pt]
\tikzstyle{block} = [rectangle, draw, fill=blue!20, 
    text width=6em, text centered, rounded corners, minimum height=3em]
\tikzstyle{blockone} = [rectangle, draw, fill=blue!20, 
    text width=6em, text centered, rounded corners, minimum height=6em]    
\tikzstyle{line} = [draw, -latex']
    
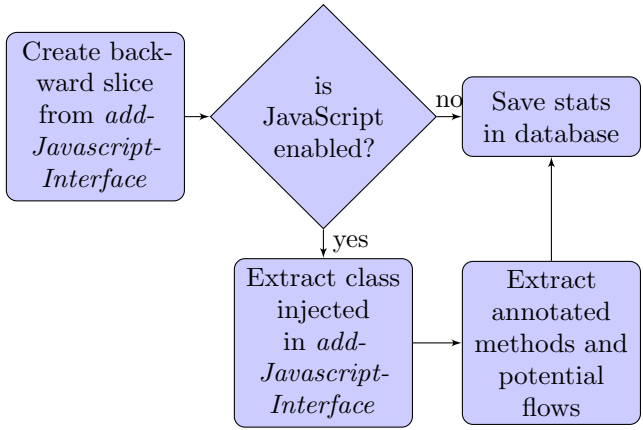
\begin{figure}[t!]   
\centering
\begin{tikzpicture}[node distance = 3cm, auto]
 \node [blockone] (init) {Create backward slice from \textit{addJavascriptInterface}};
 \node [decision, right of=init] (decide) {is \JS enabled?};
 \node [block, below of=decide, node distance=3cm] (yes1) {Extract class injected in \textit{addJavascriptInterface}};
 \node [block, right of=yes1, node distance=3cm] (final) {Extract annotated methods and potential flows};
 \node [block, right of=decide, node distance=3cm] (no1) {Save stats in database};

\path [line] (init) -- (decide);
\path [line] (decide) -- node {yes} (yes1);
\path [line] (yes1) -- (final);
\path [line] (decide) -- node {no} (no1);
\path [line] (final) -- (no1);
\end{tikzpicture}
\caption{The workflow of IFCAnalyzer}
\label{fig:jsworkflow}
\end{figure}

\subsection{ResourceExtractor}\label{ResourceExtractor}
The \emph{loadUrl} methods load a specified URL given as string parameter (cf. Listing~\ref{listing:MainActivity1}, line~\ref{lst:line1_11}). If the input string starts with \emph{"javascript:"}, the string is executed as \JS code (cf. Listing~\ref{listing:MainActivity1}, line~\ref{lst:line1_9}). The aim of this module is to extract the URLs and \JS code passed to the \emph{loadUrl} methods and \emph{evaluateJavascript}. We create an intra-procedural backward slice from each of these method calls, extract the string parameters, and store them alongside their originating class' name. As string parameters are often constructed via various \emph{String} operations, (e.g., using the \emph{StringBuilder} class to concatenate strings), we extended LUDroid with domain knowledge on the semantics of the Java \emph{String} class. When LUDroid detects the Smali signature of a \emph{String} or \emph{StringBuilder} method it applies partial evaluation based on the method's semantics to statically infer the result created by the string manipulation. However, at the time of this writing we do not support the automatic resolution of complex string operations based on manipulating elements of a character array. Future work may extend the support for such operations and thus simple obfuscations. For now we concentrate on the features of the string parameters that can be extracted with reasonable effort, in order to gain timely insights on what is required for static analysis of hybrid apps. The output of this module is fed to the \emph{UrlAnalyzer} and \emph{JSAnalyzer} modules to interpret and categorize the URLs and \JS.

%% file: sections/methodology-urlanalyzer.tex

\emph{URLAnalyzer} has two functions: (1) it checks the validity of a URL, and (2)
extracts its essential features. \emph{URLAnalyzer} parses
the URLs received from \emph{ResourceExtractor} and extracts the
following set of features:
\begin{compactitem}
\item \textbf{Protocol} - The application layer protocol of the URL, e.g., HTTP. 
\item \textbf{Host} - This can either be a fully qualified domain or an IP address of the corresponding host. 
\item \textbf{Port} - The port of the host the request is sent to (optional).
\item \textbf{Path} - The path on the host the request is sent to (optional). Such a path can for example be specified for HTTP or FTP URLs, but also for local file URLs.
\item \textbf{Search} The search part of HTTP URLs (optional). This is the remainder of a HTTP URL after the path, e.g., ''?x=5\&y=9''.
\item \textbf{Fragment} The fragment is an optional part of the URL that is placed at the end of the URL and separated by \#.
\end{compactitem}

RFC 3986~\cite{berners2004uniform} defines the specification of URLs in
augmented Backus-Naur form. \emph{URLAnalyzer} validates each provided URL
against this definition in order to detect malformed URLs. For every URL,
\emph{URLAnalyzer} either confirms that the URL is syntactically correct, or prints a
detailed message why the URL is malformed. 

We also categorize the URLs created by third-party libraries (SDKs). These libraries use \emph{loadUrl} to provide the intended functionality to app developers, e.g., the Facebook SDK for Android provides Facebook's authentication service. 
Finally,
\emph{URLAnalyzer} creates a database containing the analysis results, so they can be reported by
the \emph{Reporting} module. 

With respect to the above features we answer the following questions:
\begin{compactitem}
\item \textbf{RQ2.1:} \emph{What is the distribution of protocols used in loadURL?}
\item \textbf{RQ2.2:} \emph{What percentage of URLs point to files on the device that are assumed to be trusted as they were bundled with the application?}
\item \textbf{RQ2.3:} \emph{What is the distribution of hosts? Do host hotspots exist, i.e., hosts that requests are being sent to from many different applications?}
\item \textbf{RQ2.4:} \emph{What is the distribution of resource access within one host discriminated by its path?}
\item \textbf{RQ2.5:} \emph{What percentage of URLs leverage unencrypted network communication e.g., HTTP, FTP?}
\item \textbf{RQ2.6:} \emph{Which of the external SDKs cannot be identified and are considered untrusted?}
\end{compactitem}

%% file: sections/methodology-jsanalyzer.tex

\emph{JSAnalyzer} summarizes patterns found in \JS passed to both variants of \emph{loadUrl} and \emph{evaluateJavascript} (i.e.~the strings constructed by \emph{ResourceExtractor}).
The results are stored in a database for further manual analysis with respect to the features described in the
sequel. The components of JSAnalyzer primarily consist of
scripts for automation and reporting. 
\subsubsection{Information Flow from \JS to Android} The Android SDK permits
to annotate \emph{setter} methods with \emph{JavascriptInterface}.
Transmitting the results from a web-based/\JS method to the Android object
supports reuse of existing web-based components in Android. It creates an
information flow from the external web application to the Android app. In this
paper, we identify use-cases of this behavior.
\subsubsection{Obfuscated and Unsecured Code} Many third-party libraries
employ code obfuscation to protect their intellectual property. 
At the same time it is possible to inject remote third-party libraries in \JS using unsecured
protocols such as HTTP. 
In this work we identify patterns in which 
external libraries are obfuscated or included insecurely.
\subsubsection{Passing of Sensitive Information to Third Parties} Many apps pass device specific
information to third-party libraries. This sensitive information is leveraged by third-party
libraries to enhance their services, such as targeted advertising. However, it can be detrimental to user privacy. In
this work, we identify cases of passing sensitive information to third-parties.

In particular, we answer the following questions 
\begin{compactitem}
    \item \textbf{RQ3.1}: \emph{How frequent is  third-party script injection used in \JS passed to \emph{loadUrl} or \emph{evaluateJavascript}?}
    \item \textbf{RQ3.2}: \emph{Is there non-trivial information flow from \JS to Android?}
    \item \textbf{RQ3.3}: \emph{Do third-party libraries leverage obfuscation for their \JS code?}
\end{compactitem}

%% file: sections/dataset.tex

We curate four different datasets containing both benign and malware apps. Our rationale to study both is the following: (1) the analysis of benign apps conveys information on how developers use Android-Web hybridization in practice, and (2) analysis on malware provides relevant insights on the use of the feature for malicious purposes. The former exhibits patterns which can be dangerous and potentially exploited. The latter demonstrates that these exploits have been exploited in practice. 

We select the benign apps from the Google Play Store and malware from the Argus AMD dataset~\cite{amdMalwares}. To obtain the benign apps, we crawled the Google Play Store based on the criteria  (\ref{para:benign-apps}) and (\ref{para:top-downloaded}) below. The Argus AMD dataset is the state of the art malware dataset available to the best of our knowledge.  
In what follows, we describe the datasets chosen for our study.

\begin{enumerate}[label=\Alph*]
\item \label{para:benign-apps} \textbf{Benign apps.} In this dataset we randomly chose 7,500 apps published on the Google Play Store between 2015-2019.
\item \label{para:top-downloaded} \textbf{Frequently Used Apps.} We selected a total of 144 of the top downloaded apps from 11 app categories on the Google Play Store. These categories are Banking, Business, Education, Entertainment, Health, Music, News, Online payments, Shopping, Social, and Travel, and are the categories of apps that were among the highest downloaded on Play Store. In addition, we also selected the top downloaded apps (referred as top apps) across all categories and apps that yield the most revenue (referred as top grossing), consisting of 52 apps. In total, we curate 13 categories of apps. Table~\ref{table:js-apps-categories} lists these categories together with the apps in each category as of Dec 2019, when they were downloaded.
\item \label{para:malware} \textbf{Malware:}  The Argus AMD dataset contains 24,553 samples from 71 different families of malware. The number of malware in each family range from 4 to 7843. To have a representative from each malware family, we chose at least one from each family, but choosing up to approx. $20\%$ of the malwares from each family. In total, we choose 1000 malware for the study.
\end{enumerate}

%% file: sections/evaluation.tex

All experiments were performed on a MacBook Pro with a 2.9 GHz Intel Core i7 processor, 16 GB DDR3 RAM, and MacOS Mojave 10.14.1 installed. We used a JVM version 1.8 with 4 GB maximum heap size. In the following, we provide the inferences from our evaluation for each of the aforementioned datasets.

We evaluated \emph{LuDroid} on more than 7,500 random applications from the Google Play Store to understand hybrid apps' communication patterns in the wild.

\subsection{IFC from Android to \JS in benign apps}
\input{sections/evaluation-HIF.tex}

\subsection{URL statistics in benign apps}
\input{sections/evaluation-URL.tex}

\subsection{Vulnerability Case Study: Unprotected URLs}
\input{sections/evaluation-VulnCaseStudy.tex}



%% file: sections/evaluation-HIF.tex

\pgfplotsset{width=6cm,compat=1.9}
\begin{figure}[tb]
\begin{tikzpicture}
  \begin{axis}[
    xbar,
    y axis line style = { opacity = 0 },
    axis x line       = none,
    tickwidth         = 0pt,
    enlarge y limits  = 0.2,
    enlarge x limits  = 0.02,
    anchor=north,
    symbolic y coords = {Sensitive Flows, Injects Class, JS Enabled, Uses WebView},
    nodes near coords,
  ]
  \addplot  coordinates { (1330,Sensitive Flows)         (2256,Injects Class)
                         (4469,JS Enabled)  (5083,Uses WebView) };
                           
 \end{axis}
\end{tikzpicture}
\caption{Hybrid API usage (Over 7500 apps)}
\label{fig:Picture1}
\end{figure}
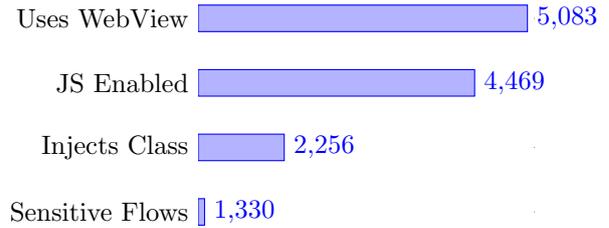 

\begin{table*}[tb]
\caption{Top ten app categories with type of information shared from Android to \JS}
\centering
\begin{tabular}{ll} 
 \toprule
  \textbf{App category} &  \textbf{Type of information} \\
 \midrule
 Social & Cookies, File system\\
\hline
Entertainment & Account Information, File system, Network Information, Location \\
\hline
Music \& Audio & Account Information, File system, Network Information \\
\hline
LifeStyle & Activity Information, Application level navigation affordances, Locale \\
\hline
Board Games & Date and Time, Location, Network Information\\
\hline
Communication & Activity Information, File system, Location, Network Information\\
\hline
Personalization & Activity Information, Account Information, File system,  Location \\
\hline
Books \& Reference & File System, Location, Network Information \\
\hline
Puzzle & File System, Internal Memory Information, Location  \\
\hline
Productivity & File System, Internal Memory Information, Network Information \\
\bottomrule
\end{tabular}
\label{table:mapping}
\end{table*}

\textbf{RQ1.1:} \textit{How pervasive is information flow from Android to \JS?}
Figure~\ref{fig:Picture1} provides the distribution of apps based on various characteristics of hybrid communication. 68\% out of 7,500 apps use WebView at least once, i.e. are hybrid apps, which is a significantly high percentage. As \JS is not enabled by default, 87.9\% of hybrid apps enable \JS while the remaining 12.1\% use WebView solely for static webpages. Half of the components enabling \JS establish an interface to \JS via \emph{addJavascriptInterface} and bridge an Android object to \JS. Therefore, 30\% of the apps used in our dataset and 43\% of the hybrid apps transfer information from Android to \JS. Table~\ref{table:mapping} presents the top ten app categories and the corresponding types of information shared with \JS. Note that in this work we do not investigate what happens to this data on the \JS side, i.e., whether it actually leaks to some untrusted entity. The focus instead is to identify scenarios in the wild that need to be taken into consideration when attempting to design an analysis for hybrid apps.

\begin{table*}[tb]
  \centering
\caption{App components with the shared sensitive information (from Android to \JS)}
	\small
\begin{tabular}{lllp{0.25\textwidth}} 
 \toprule
  \textbf{App Name} & \textbf{Category} & \textbf{Component Name} & \textbf{Information shared} \\
 \midrule
 Instagram & Social & BrowserLiteFragment & Cookies\\
\hline
 TASKA AR MARYAM & Entertainment & Map26330 & Location (GPS)\\
\hline
Classical Radio & Musik \& Audio & MraidView & External storage file system access, Network Information \\
\hline
BLive & Lifestyle & LegalTermsNewFragment & Location \\
\hline
Cat Dog Toe & Board Games & appbrain.a.be &  Location, Network Information \\ 
\hline
N.s.t. A-Tech & Communication & ax & Location, Network Information \\
\hline
Pirate ship GO Keyboard & Personalization & BannerAd & Device ID, Device's Account information, Locale\\
\hline
IQRA QURAN &  Books \& Reference & Map26330 & Location\\
\hline
Logic Traces & Puzzle & SupersonicWebView & Location\\
\hline
FLIR Tools Mobile & Productivity & LoginWebActivity & Network Information \\
\bottomrule
\end{tabular}
\label{table:sensitive}
\end{table*}

\textbf{RQ1.2:} \textit{Do these information flows include sensitive information?}
18\% of the total apps in our dataset share sensitive information from Android to \JS. LUDroid finds 6375 sensitive information flows from Android to \JS: Only 18\% of these flow to URLs located inside the app, i.e., using the \textit{file} protocol. Note that the inclusion of \JS code into an app does not guarantee its trustworthiness, as third party code is regularly included into apps. Thus 82\% (or more) of the sensitive information flows could leak to potentially untrusted code. Table~\ref{table:sensitive} presents 10 randomly selected apps\footnote{Due to the size limitation we could not publish the entire list.} for each category mentioned in Table~\ref{table:mapping} along with their corresponding components and shared sensitive information. The majority of these flows include location information, network information and file system access. 
Starting from HTML5, various web APIs provide access to sensitive information such as the geographical location of a user. In contrast to Android's permission system where users need to approve the permissions just once (potentially in a completely different context), web users would need to approve the access each time or they can provide it for one day. It appears that this might be one of the reason that developers prefer to propagate sensitive information from Android to the Web, at the expense of users' privacy.

%% file: sections/evaluation-URL.tex

\label{sec:url-eval}

\newcommand{\varResolvedURLs}{3075}

\newcommand{\varNumberURLsToNetworkResource}{535}
\newcommand{\varNumberIndividualPaths}{147}

\newcommand{\varTotalNumberUnencryptedNetworkCalls}{365}
\newcommand{\varPercentageUnencryptedNetworkCalls}{11.87\%} 

\newcommand{\varPercentageURLsSyntacticallyInCorrect}{0.58\%} 
\newcommand{\varPercentageURLsToLocalFile}{40.81\%} 
\newcommand{\varAboutBlankPercentage}{41.24\%} 

\newcommand{\varPercentagePortSpecified}{1.68\%}
\newcommand{\varPercentageArgumentPatternSpecified}{20.37\%}

\newcommand{\varNumberURLsDynamicallyCreated}{4980}

\newcommand{\varPercentageMonetizationAnalytics}{18.04\%}
\newcommand{\varPercentageUsingUntrustedSDK}{13.89\%}
\newcommand{\varPercentageUsingCloudAppDevelopment}{4.17\%}

\begin{figure}[tb]
\centering
\begin{tikzpicture}[scale=0.8]
    \pie[text=pin]{6/https, 12/http, 41/file, 41/about:blank}
\end{tikzpicture}
\caption{Distribution of protocols (rounded values for clarity)}
\label{fig:protocol-distribution}
\end{figure}
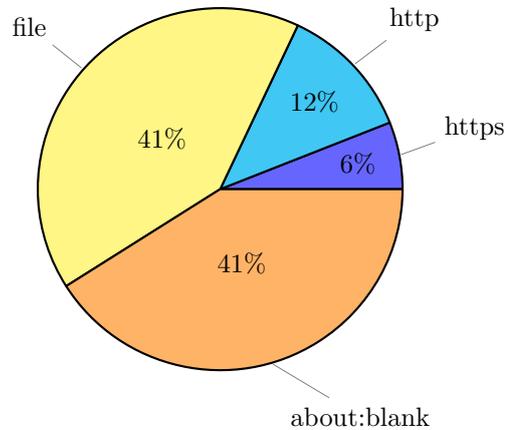

LUDroid resolved \varResolvedURLs{} distinct URLs. In addition it found \varNumberURLsDynamicallyCreated{} URLs dynamically created using SDKs. Figure \ref{fig:protocol-distribution} shows the distribution of protocols in the resolved URLs passed to the \emph{loadURL} method \textbf{(RQ2.1)}.
\varPercentageURLsToLocalFile{} of the URLs use the \emph{file} protocol pointing to the device's (trusted) local files, while the remaining point to external (potentially trusted) hosts \textbf{(RQ2.2)}. 
Naturally, developers have more control
over these offline local files. While this is good for trusted entities, malicious entities could easily launch phishing attacks by designing offline pages that look similar to trusted web pages. Only good user practices can prevent these attacks from happening: Ideally, APKs should not be downloaded from other sources than the official Play Store. Additionally, users should properly verify app metadata and permissions. 

 As local web pages come bundled with the APK files, they can be taken into account during analysis. However, an analysis might need to consider several security aspects such as identifying phishing attacks, discovering privacy leaks, or finding keyloggers.

\begin{lstlisting}[language=Java, belowskip=-0.8 \baselineskip, caption=WebViewActivity class in com.zipperlockscreenyellow (manually translated to Java and simplified), label={listing:zipperl-lock-code-java}, float=tb, escapeinside={*}{*}]
webView.removeAllViews();
webView.clearHistory();
webView.clearCache();
webView.loadURL("about:blank");
\end{lstlisting}
In addition to local file URLs, we discovered that in \varAboutBlankPercentage{} of the resolved cases the URL argument was \emph{''about:blank''}, which displays an empty page. According to Android's WebView~\cite{webview} documentation \emph{about:blank} should be used to ``reliably reset the view state and release page resources''. As an example, we discovered \emph{about:blank} in the WebViewActivity class of the app \emph{com.zipperlockscreenyellow}. In this class the method \emph{killWebView} releases the view's resources (see Listing~\ref{listing:zipperl-lock-code-java}). After clearing the history and the cache this method opens a blank page in the WebView.

\begin{figure}[t]
\begin{tikzpicture}
  \begin{axis}[
    xbar,
    height=9cm,
    y axis line style = { opacity = 0 },
    axis x line       = none,
    tickwidth         = 0pt,
    enlarge y limits  = 0.14,
    enlarge x limits  = 0.02,
    anchor=north,
    scaled y ticks=false,
     symbolic y coords = {Social,App Monetization,Web Services,Untrusted/Unknown,Online App Generator,Outsourcing,Mobile Development,ECommerce},
    nodes near coords, nodes near coords align={horizontal},
  ]
  \addplot  coordinates { (27.77,Social)    (20.94,App Monetization)
                         (16.46,Web Services)  (13.89,Untrusted/Unknown)
                         (8.64,Online App Generator) (3.44,Outsourcing)
			(1.26,Mobile Development) (1.15,ECommerce)};
                          
 \end{axis}
\end{tikzpicture}
\caption{Distribution of SDK usage in apps by categories (Top eight)}
\label{figure:sdk-usage-categories}
\end{figure}
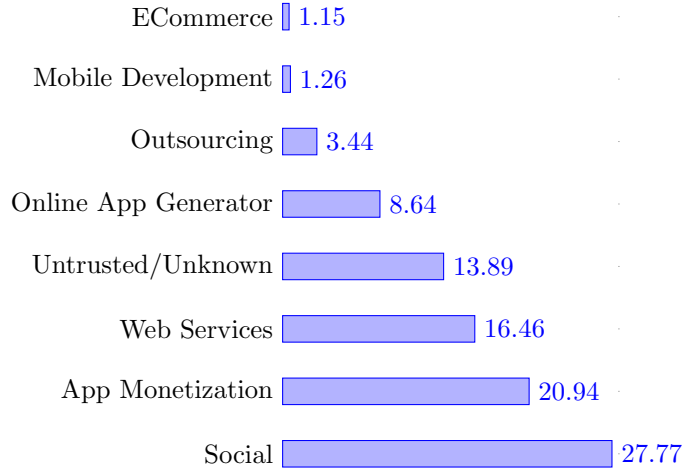

\begin{table*}[tb]
	 \caption{Selected list of Top-8 SDK hosts, its app share, and a common use case found in the top SDK categories}
    \centering
    \begin{tabular}{llrl}
        \toprule
        \textbf{Category} & \textbf{Host}  & \textbf{Percent}  & \textbf{Common Use Case}\\
        \midrule
        Social Networking &Facebook & 20.32 & Authentication \\
        \hline
        App monetization & Vungle & 1.75 & Monetize Apps by targeted advertising \\ 
        \hline
        Web Services & Google   & 10.58 & Authentication \\
        \hline
        Online App Generator & SeattleClouds & 5.99 &  Unknown (obfuscated) \\
        \hline
        Outsourcing & biznessapps & 1.89 & Unknown (obfuscated) \\
        \hline
        Mobile Development & PhoneGap & 1.52 & Platform-independent development \\
        \hline
        E-Commerce & Amazon & 1.1 & Sales \\
        \hline
        Others & Ons & 0.8 & Rendering ebooks\\
        \bottomrule
    \end{tabular}
    \label{table:sdk-usage}
\end{table*}

Considering network URLs, the most-frequently loaded hosts per category in the analyzed apps are listed in Table~\ref{table:sdk-usage}. We found that Facebook and Google SDKs are widely used in apps, primarily for authentication purposes. In addition app monetization and customer analytics SDKs are found in \varPercentageMonetizationAnalytics{} of the apps (cf.~Figure~\ref{figure:sdk-usage-categories}). Figure~\ref{figure:sdk-usage-categories} displays all host categories sorted by their share \textbf{(RQ2.3)}. We found that a majority of the analyzed apps use social networking SDKs or app monetization SDKs.

Mobile application development frameworks such as Cordova or PhoneGap  allow developers to use HTML/CSS and \JS to develop mobile apps. These libraries primarily  use bridge communication between native Android and web technologies~\cite{phonegap2010}. In our study we found that $1.26\%$ of the apps use these frameworks for mobile application development (referred in Figure~\ref{figure:sdk-usage-categories} as Mobile Development).

The \varNumberURLsToNetworkResource{} URLs that point to a network resource only reference \varNumberIndividualPaths{} distinct paths. This indicates that in many cases identical host and path combinations were requested by multiple apps \textbf{(RQ2.4)}. In approximately \varPercentagePortSpecified{} of the external URLs the host's port was specified. Additionally, \varPercentageArgumentPatternSpecified{} of the URLs (HTTP/HTTPS) specify an argument pattern.

\begin{table}[tb]
\centering
\caption{Five out of 365 loadURL calls using the insecure HTTP protocol}

\begin{tabular}{m{0.45\columnwidth} m{0.45\columnwidth}} 
\toprule
\textbf{Package Name} & \textbf{URL}\\ 
\midrule
com.JLWebSale20\_11 & \url{http://www.dhcomms.com/applications/dh/cps/google/main_agreepage01.html}\\ 
\hline
net.pinterac.leapersheep.   main & \url{http://pinterac.net/dev/leapersheep/index.php?viewall=1}\\ 
\hline
com.quietgrowth.qgdroid & \url{http://docs.google.com/gview?embedded=true&url=http://www.rblbank.com/pdfs/CreditCard/fun-card-offer-terms.pdf}\\ 
\hline
net.lokanta.restoran.   arsiv\-trkmutfagi & \url{http://images.yemeksepetim.com/App_Themes/static-pages/terms-of-use/mastercard/mobile.htm}\\ 
\hline
com.cosway.taiwan02 & \url{http://ecosway.himobi.tw}\\ 
\bottomrule
\end{tabular}
\label{table:http-urls}
\end{table}

While evaluating URLs we gained several relevant security insights. We found  that \varPercentageUnencryptedNetworkCalls{} of the calls to \emph{loadUrl} resulted in unencrypted network traffic, making a total number of \varTotalNumberUnencryptedNetworkCalls{} communications. Table~\ref{table:http-urls} shows five examples of unencrypted HTTP URLs together with the packages in the corresponding app \textbf{(RQ2.5)}. The usage of unencrypted protocols with \emph{loadURL} may result in eavesdropping and phishing vulnerabilities. We demonstrate how to exploit such a
vulnerability in section~\ref{sec:VulnCaseStudy}. Another security threat is caused by untrusted SDKs using \emph{loadURL}. We found a total
of \varPercentageUsingUntrustedSDK{} apps use untrusted SDKs \textbf{(RQ2.6)}. However, in this context untrusted may or may not refer to a malicious SDK. It is non-trivial to classify untrusted SDKs as malicious due to absence of common patterns in these SDKs. Therefore, we take a conservative approach where an SDK is untrusted if there is no public information available on the web. 
Clearly, security testing of untrusted SDKs is imperative to ensure the integrity of one's code. However, many programmers include desired functionalities into their projects without considering the security implications.

Another interesting observation is the usage of online app development platforms. These development platforms allow users to build an application with
minimal technical effort and programming background. From the collected data, we found using manual inspection that approximately 8.64\% (cf.~Figure~\ref{figure:sdk-usage-categories}) of the apps use an online app generation platform. 
A potential threat to these applications is that the developer/app provider using the online app generators neither has the knowledge about the internal details of these apps nor do they perform rigorous testing. A recent study on these online app generators (OAG) found serious vulnerabilities for various OAG providers~\cite{Sacha2018OnlineApp}. 
Again, programmers should not rely blindly on the quality of external tools and perform additional validation of the resulting app's security properties. Unfortunately, OAGs are particularly intriguing to developers with low technical expertise, so the creators of these platforms have a responsibility.

We discovered 18 instances (see e.g.~URL for \emph{com.quietgrowth.qgdroid} in Table~\ref{table:http-urls}) where a call to \emph{loadURL} was used to display a PDF via Google Docs, which is considered a misuse of WebView. To deliver content such files to users the WebView documentation recommends to invoke a browser through an \emph{Intent} instead of using a WebView~\cite{google2018}. It appears that developers prefer users to stay inside the app for viewing documentation, and thus rather use \emph{WebView} to accomplish this task.

%% file: sections/evaluation-VulnCaseStudy.tex

\label{sec:VulnCaseStudy}
As described in section \ref{sec:url-eval}, \emph{URLAnalyzer} determines whether a URL passed to \emph{loadURL} is unprotected, i.e.~whether it points to a network resource and is not protected by any cryptographic means (e.g. TLS). In our evaluation we discovered \varTotalNumberUnencryptedNetworkCalls{} calls to \emph{loadURL} with unprotected URLs, all of which connect via HTTP.

The \emph{loadURL} method embeds a web page into the Android application. When using an unprotected URL for \emph{loadURL} an attacker can read the requested webpage, and even more severe, manipulate the server's response that is to be displayed to the user. This is particularly critical as an attacker-controlled webpage is then being displayed in the context of a trusted application. The user may be oblivious to the difference between content displayed in a WebView and content displayed in other UI components, as WebViews are designed to seamlessly integrate into the native UI components. Depending on the concrete application and the placement of the vulnerable WebView in the native UI, various attack scenarios are possible. One attack scenarios is a phishing attack where a malicious login page is displayed to the user within the app. As the app is trusted by its users, they are likely to enter their credentials on the phishing page.

\paragraph*{Case Study: EndingScene app}
To demonstrate the described attack, we randomly chose one vulnerable application, EndingScene (v 1.2\footnote{md5: 7516ddd1bc9d056032ac3173e71251b0}), a video material promotion app. Immediately in the initial activity, this app loads a webpage and displays it to the user. This scenario is ideal for an attacker, as every user will be presented this initial front page, and the activity consists of nothing else but the front page. In addition, it is very plausible to ask for some type of credential on this front page.

We implemented a network attack using \emph{mitmproxy}~\cite{mitmproxy}, a HTTP proxy that can save and manipulate inflowing traffic. We developed a small Python script for use in \emph{mitmproxy}. It substitutes the server's response to the front page request with a self-written malicious login page, which sends the entered credentials to an attacker.

\begin{figure*}
\centering
\includegraphics[scale=0.1]{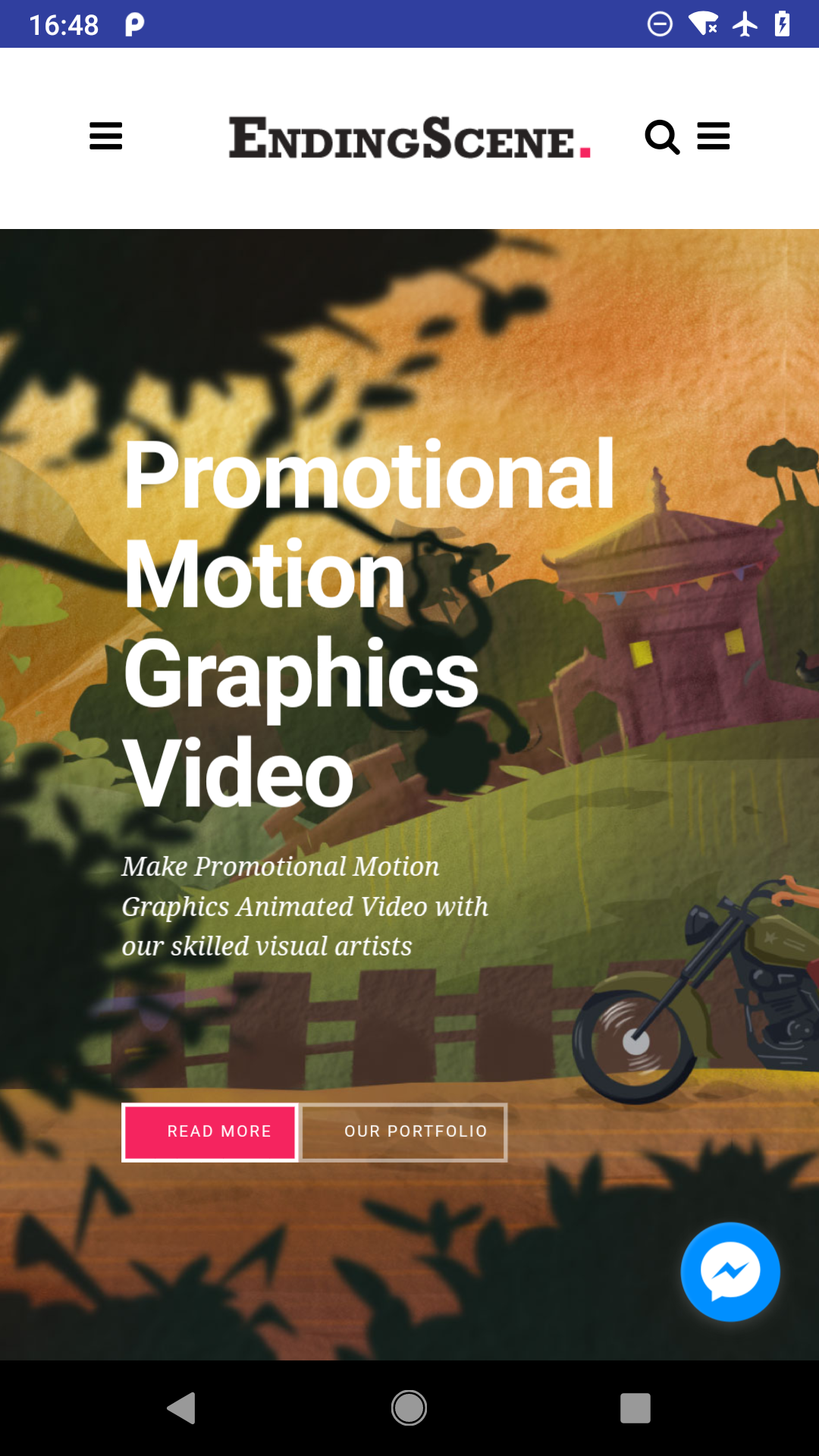}
\hspace{1cm}
\includegraphics[scale=0.1]{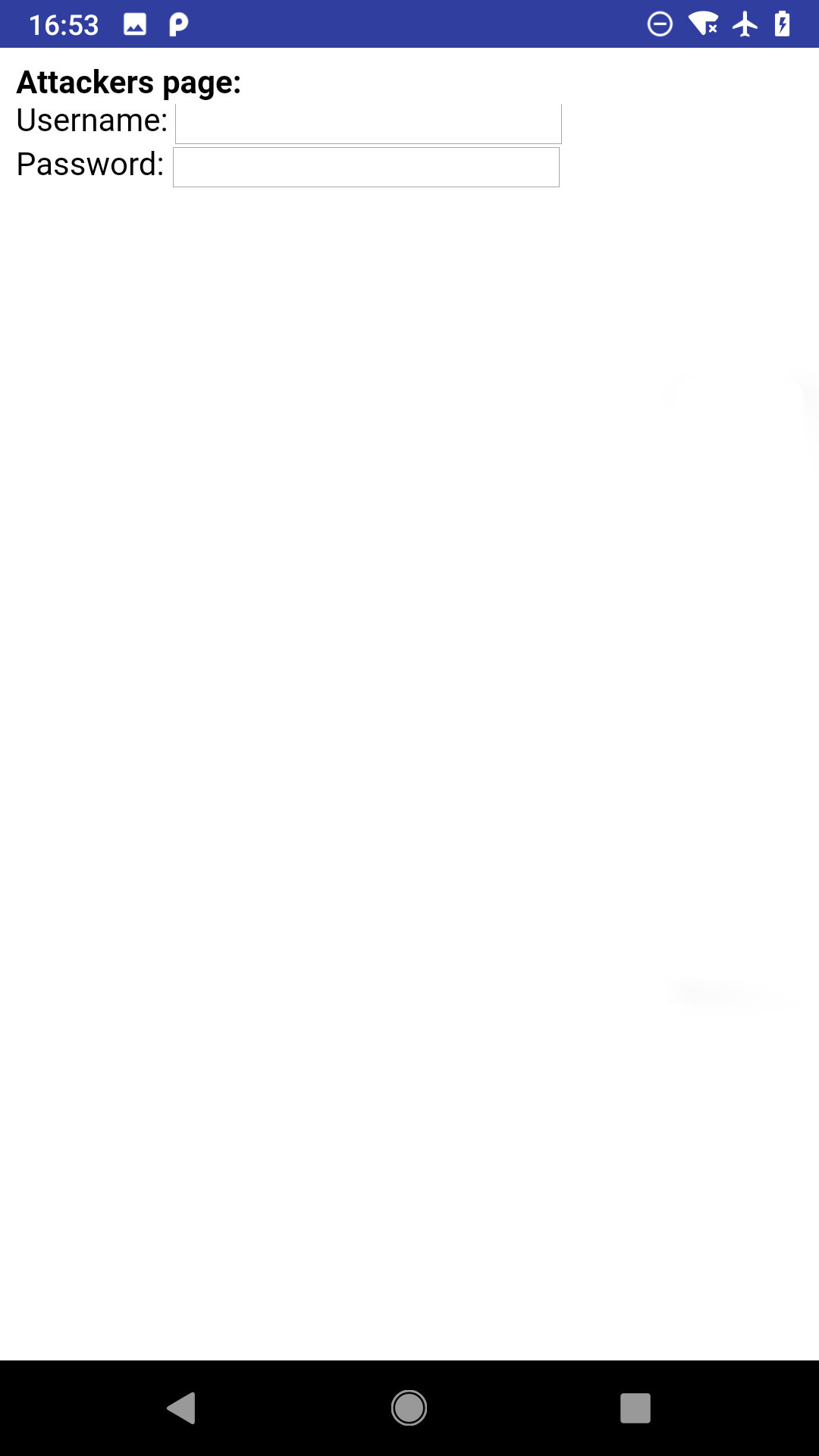}
\caption{The EndingScene app without and with being attacked (In a realistic attack the phishing web page may be easily copied from the original web page)}
\label{fig:EndingScene-screenshots}
\end{figure*}

Figure~\ref{fig:EndingScene-screenshots} depicts the successful exploitation of the EndingScene app when using our proxy. The left-hand side shows the regular front page of EndingScene while the right-hand side displays the phishing page when the network traffic is being attacked. Evidently, a malevolent entity would create a much more convincing phishing page, our page is for illustration purposes only, to make the attack obvious. This type of attack in general is not new, however, related work~\cite{pokharel2017can,he2014security,fahl2012eve} has not detected them in the context of hybrid apps, which may lead to novel attack vectors.

As the described vulnerability is caused by the lack of encryption and signatures, it may be avoided using the transport layer security (TLS) versions of the protocol (e.g. HTTPS, FTPS). Additionally, it is recommendable to make use of certificate pinning in order to prevent threats from corrupted certification authorities.

%% file: sections/evaluation-JS.tex
In what follows we present initial insights on \JS code that LUDroid identified to be passed to \emph{bridge methods}, i.e., \emph{loadUrl} or \emph{evaluateJavaScript}. 

\def\totalFreqJS {73~}

\input{sections/appendix-js.tex}

\subsection{Frequent \JS Code in Benign Apps \label{subsection:frequent-js}}

Based on our manual analysis for benign apps, we classify the \JS code into two categories: (1) involving event-driven functionality (using the interface \emph{Event}), and (2) modifying the Document Object Model~(DOM) without event-driven functionality. We found that \(64\%\) trigger event-driven functionality while  \(31\%\) modify the DOM only. We were unable to resolve $5\%$ of the \JS strings owing to the current limitations of \emph{IFCAnalyzer}. We identified a set of \totalFreqJS distinct \JS code clones (of type~2~\cite{codeClones}, i.e. syntactically identical copy where identifiers might have been replaced\footnote{As we only have access to decompiled binaries we are lacking the original identifier names in most cases.}) passed to \emph{bridge methods} in all investigated apps. Given this low number in relation to the total number, it was not surprising to identify that most of these originate from third-party libraries. Interestingly, the set of code executed using \emph{evaluateJavaScript} is a subset of those executed via \emph{loadURL}, therefore, we will restrict ourselves to the discussion of the latter in the sequel. 

Table~\ref{table:top-4-js-code} lists the four most frequently used \JS code fragments and their usage. The most commonly used JavaScript code snippet in our dataset originates from the Facebook SDK. More than 32\% of apps, in our dataset, have used it at least once. This code snippet provides the functionality to authenticate with Facebook on the web in case the Facebook mobile app is not present in the user's device. The second most common JavaScript code snippet exhibits a peculiar case. In this snippet, a DOM element modifies an Android object. We describe further details in Case Study~\ref{paragraph:case-study-js-android}. Similar functionality is manifested by the fourth JavaScript code snippet. In this case, a JavaScript code snippet, originated from an external SDK, modifies an Android object by invoking the setter method.  In particular, both cases manifest an interesting scenario where JavaScript modifies Android objects. Finally, the program in the third row originates from the advertisement  library Vungle. Details regarding this case are described in Case Study~\ref{sec:vungle}.

In what follows, we illustrate the four most interesting cases relevant to understand the developers' intentions.

\begin{lstlisting}[language=JavaScript, aboveskip=0.1\baselineskip, belowskip=-0.1\baselineskip, caption={Dynamic Script Loading in loadUrl}, label={listing:dynamic-script},float=tb]
javascript: (function() { var script=document.createElement('script');
script.type='text/JavaScript';
(*\label{listing:dynamic-script:3}*)script.src='http://admarvel.s3.amazonaws.com/   js/admarvel_mraid_v2_ complete.js';
document.getElementsByTagName('head')               .item(0).appendChild(script);})()
\end{lstlisting}
\subsubsection{Case Study: Third-party script injection in loadUrl}
We identify the case of a third-party script injection that occurs in 3 of the \totalFreqJS codes identified. One example of such a script is displayed in Listing~\ref{listing:dynamic-script}. Similar instances are present in $8.33\%$ \textbf{(RQ3.1)} of the analyzed apps. The script loads third-party \JS code by injecting a script tag into the header of the displayed webpage, resulting in a modification of the global state of the page. In this example, the developers used an unsecured protocol (\emph{HTTP}, cf.~Line~\ref{listing:dynamic-script:3}, Listing~\ref{listing:dynamic-script}). This scenario makes the webpage and thus the whole Android app susceptible to a man-in-the-middle (MITM) attack, where an attacker can intercept the connection and replace the script loaded from \verb|script.src| with malicious \JS. However, the user trusts the app and is completely oblivious to the script being downloaded and that it might be replaced, which violates the integrity of the app. This attack can be implemented analogously to the attack described in Section~\ref{sec:VulnCaseStudy}, where the login page was substituted by a malicious page.

\subsubsection{Case Study: Information flow from \JS to Android}\label{paragraph:case-study-js-android}

\begin{lstlisting}[language=JavaScript, aboveskip=0.1\baselineskip, belowskip=-0.1\baselineskip, caption={Modifying the bridged Android object named SynchJS}, label={listing:writing-android-objects}, float=tb]
(*\label{listing:writing-android-objects:1}*)javascript:window.SynchJS.setValue((function(){
(*\label{listing:writing-android-objects:3}*)    try{
(*\label{listing:writing-android-objects:4}*)        return JSON.parse(Sponsorpay.MBE            .SDKInterface.do_getOffer()).uses_tpn;
(*\label{listing:writing-android-objects:5}*)    }catch(js_eval_err){
(*\label{listing:writing-android-objects:6}*)        return false;
(*\label{listing:writing-android-objects:7}*)}})());
\end{lstlisting}
\def\percentAboutBlankJs {4.17\%}
\def\dataflowJsToAndroid {8.7\%}
Contrary to common intuition we identified interesting cases of information flow from \JS to
Android in $\dataflowJsToAndroid$ \textbf{(RQ3.2)} of the investigated apps.
Listings~\ref{listing:writing-android-objects} and \ref{listing:modify-android-1}
display examples of this behavior. 

\begin{lstlisting}[caption={Excerpt of source code for SynchJS object in Listing~\ref{listing:writing-android-objects}. Source: Github~\cite{github2019}},float=tb,label={listing:android-setvalue},escapeinside={(*}{*)},belowskip=-\baselineskip, aboveskip=-0.1\baselineskip, float=tb, language={Java}, breaklines=true]
public class SynchronousJavascriptInterface {
    // JavaScript interface name for adding to web view
    private final String interfaceName = "SynchJS";
    private CountDownLatch latch; // Countdown latch to wait for result
    private String returnValue; // Return value to wait for
(*\label{listing:android-setvalue:getJSValue}*)    public String getJSValue(WebView webView, String expression) {
      latch = new CountDownLatch(1);
      String code = "javascript:window." + interfaceName + ".setValue((function(){try{return " + expression + "+\"\";}catch(js_eval_err){return '';}})());";
      webView.loadUrl(code);
      try { // Set a 1 second timeout in case there's an error
(*\label{listing:android-setvalue:await}*)        latch.await(1, TimeUnit.SECONDS);
(*\label{listing:android-setvalue:return}*)        return returnValue;
      } catch [...] return null; }
    // Receives the value from the JavaScript.
    public void setValue(String value) {
(*\label{listing:android-setvalue:131}*)      returnValue = value;
      try { latch.countDown(); } catch (Exception e) {} 
    }
}
\end{lstlisting}

Listing~\ref{listing:writing-android-objects} is particularly interesting as it leverages a  synchronous communication channel from Android to \JS and back. In Listing~\ref{listing:writing-android-objects}, a method \emph{setValue()} is invoked on a bridged object \emph{SynchJS}. The method \emph{setValue()} is a setter method defined in the class \emph{SynchronousJavascriptInterface} excerpted in Listing~\ref{listing:android-setvalue}. Note that the code of Listing~\ref{listing:writing-android-objects} is generated in the method \emph{getJSValue} (line~\ref{listing:android-setvalue:getJSValue}), where Android executes the parameter expression in the context of the WebView and waits (line~\ref{listing:android-setvalue:await}) for the thread evaluating the \JS code to invoke the bridged \emph{setValue} method. Line~\ref{listing:writing-android-objects:4} in Listing~\ref{listing:writing-android-objects} reads the field \emph{uses\_tpn} of an object deserialized from a third-party library method \emph{Sponsorpay.MBE.SDKInterface.do\_getOffer} and passes that value to the setter method in \emph{SynchJS}. When this method is invoked inside the WebView's thread, the field  \emph{returnValue} is changed (line~\ref{listing:android-setvalue:131} of Listing~\ref{listing:android-setvalue}). The implementation then notifies Android's UI thread via a call to \emph{latch.countDown()}, which basically implements a simple semaphore such that the waiting Android thread can continue its execution and return the value retrieved from the WebView (line~\ref{listing:android-setvalue:return}). 

\begin{lstlisting}[language=JavaScript, belowskip=-\baselineskip, aboveskip=0.1\baselineskip, caption={Information Flow from \JS to Android}, 
    label={listing:modify-android-1},float=tb]
(*\label{listing:modify-android-1:1}*)(function() {  
(*\label{listing:modify-android-1:2}*)  var metaTags=document.getElementsByTagName('meta');  
(*\label{listing:modify-android-1:3}*)  var results = [];  
(*\label{listing:modify-android-1:4}*)  for (var i = 0; i < metaTags.length; i++) {    
(*\label{listing:modify-android-1:5}*)    var property = metaTags[i].getAttribute('property');    
(*\label{listing:modify-android-1:6}*)    if (property && property.substring(0, 'al:'.length) === 'al:') {
(*\label{listing:modify-android-1:7}*)        var tag = { "property": metaTags[i].getAttribute('property') };     
(*\label{listing:modify-android-1:8}*)        if (metaTags[i].  hasAttribute('content') ) {        
(*\label{listing:modify-android-1:9}*)            tag['content'] = metaTags[i].getAttribute('content');
(*\label{listing:modify-android-1:10}*)    }      
(*\label{listing:modify-android-1:11}*)    results.push(tag); 
  } //if end
} //for end
(*\label{listing:modify-android-1:14}*)window.HTMLOUT                 .processJSON(JSON.stringify(results));
})()
\end{lstlisting}
Listing~\ref{listing:modify-android-1} writes meta-tags information of a HTML page to an Android object. Line~\ref{listing:modify-android-1:5}-\ref{listing:modify-android-1:11} construct an array of objects with properties \emph{property} and \emph{content}. This array is then converted to a string in \JS Object Notation (JSON) representation (line~\ref{listing:modify-android-1:14}) before being passed to the
\emph{processJSON()} method of the bridged object \emph{HTMLOUT}. Note that due to the fact that the \emph{processJSON} method runs in a different thread than the regular Android code~\cite{google2018} the Java Memory Model may not allow the Android code to see any changes performed to the state of the bridged (and other) objects unless some form of synchronization is being used as in Listing~\ref{listing:android-setvalue}.

\begin{lstlisting}[language=Java,label={listing:smali-dataflow-android-js},caption={Representative Java listing called from \emph{onPageFinished()} in \emph{Liberty Education App}},float=tb,escapeinside={(*}{*)}, aboveskip=0.2\baselineskip, belowskip=-\baselineskip]
public void process() {
    ....
    WebView v0 = p0.viewFinished;
    ...
    String v1 = "javascript:..."; //String from Listing (*\ref{listing:modify-android-1}*)
    v0.loadUrl(v1)
    ...
}
\end{lstlisting}
Android WebViews feature an event system that reacts to many different events in the WebView. 
The Android SDK allows developers to override the default \emph{WebView chromeless browser} window and specify their own policies
and window behavior through extending a Java interface called
\emph{WebViewClient}. Interestingly, we also identified many similar code fragments during handling of \emph{WebViewClient} events.  As an example, developers can modify the behavior when e.g. the
WebView client is closed. Our study found that many developers transfer results
from \JS to Android after the WebView client terminates by modifying the
\emph{onPageFinished()} method in the \emph{WebViewClient} interface to invoke \JS.
Listing~\ref{listing:smali-dataflow-android-js} shows an example taken from the app
\emph{Liberty Education} where the method \emph{process()} is called from method \emph{onPageFinished()}
by a series of method calls. 

Our study shows the use of  sophisticated patterns by developers for
communication from \emph{Android} to \emph{\JS} and vice-versa. 
Our study shows intricate cases of using setter methods to permit non-trivial
dataflow from \JS to Android, in some cases even using (required) synchronization.

Restricting the bidirectional communication impacts the flexibility provided by WebView to
developers. Instead static analysis techniques could be leveraged to detect and report similar insecure
data flows.  However, a simple context-insensitive static analysis on
Listing~\ref{listing:writing-android-objects} using approaches such as
HybridDroid~\cite{lee2016hybridroid}, or Bae et. al.~\cite{Ryu2019} will be unsound. The
unsoundness stems from the analyses' limitation to analyze the described callback communication methods,
thus only supporting one-way communication from Android to \JS.
A precise and sound static analysis would need to consider these non-trivial methods of callback communication that establish a two-way communication channel.

\subsubsection{Case Study: Device Information to Third-party}\label{sec:vungle}
This case study illustrates leakage of device information to third-party
libraries. Listing~\ref{listing:leaking-information} is taken from the
advertising library \emph{Vungle}. Line~\ref{listing:leaking-information:8}
removes the highlight color from each element. Therefore, the function
\emph{noTapHighlight()} makes the app susceptible to a confused-deputy attack such as
clickjacking.  It obscures user clicks, making users click on
advertisements without their knowledge. Additionally,
Line~\ref{listing:leaking-information:11} can potentially leak Vungle's
\emph{WebView} configuration object, which contains identifiable information of a device, to some web server, in this example
through the function \emph{vungleInit()}. \emph{WebView} settings contain sensitive
information about the host device that is also used by
\emph{WebViewClient}. 

\begin{lstlisting}[language=JavaScript, belowskip=-\baselineskip, caption={Leaking Sensitive Information}, label={listing:leaking-information},float=tb]
function actionClicked(m,p) { 
    var q = prompt('vungle:' + JSON.stringify({method:m, params:(p?p:null)}));
    if(q && typeof(q) === 'string') {
        return JSON.parse(q).result;
    }
};
function noTapHighlight(){
    var l=document.getElementsByTagName('*');
    for(var i=0; i<l.length; i++){
(*\label{listing:leaking-information:8}*)      l[i].stylewebkitTapHighlightColor=  'rgba(0,0,0,0)';
    }
};
noTapHighlight();
(*\label{listing:leaking-information:11}*)if (typeof vungleInit == 'function') {
    vungleInit($webviewConfig$);
}
\end{lstlisting}

\subsubsection{Case Study: Code Obfuscation in Third-Party Libraries}
This case study shows an interesting obfuscation pattern using \emph{loadUrl} to deliberately prevent program analyzers from inferring the intended functionality. Need for obfuscation arises from concerns about safeguarding the intellectual property, or from trying to hide debatable or, worse, malicious behavior. 

\begin{lstlisting}[language=JavaScript, belowskip=-\baselineskip, caption={Complex control flow via \JS}, 
    float=tb, label={listing:code-obfuscation}]
javascript:(function() { Appnext.Layout.destroy('internal'); })()
\end{lstlisting}
Appnext is an ad-library which is widely used for app monetization. 
Listing~\ref{listing:code-obfuscation} shows a code snippet found in its library code. In this code a Java object \emph{Appnext} is being bridged and used in \JS invoked from Android. This functionality could have been directly implemented in Android/Java itself. It is unclear why the programmers chose to implement it by crossing a language-bridge from Android to \JS and back, which is even more expensive as an \emph{eval} in \JS, instead of the direct invocation. We have discovered this pattern in $25\%$ \textbf{(RQ3.3)} of the apps, which makes this potential obfuscation pattern prevalent among Android apps. In this case, the anonymous function can be directly replaced by the function call. 

\begin{lstlisting}[language=Java,label={listing:obfuscation-adlocus},caption={Representative Java listing of obfuscation in Library Code-AdLocus},belowskip=0.2\baselineskip, aboveskip=0.2\baselineskip, float=tb]
package com.adlocus.adapters;
public class AdLocusAdapter {
    ...
    protected final WeakReference a;
    ...
    private WebView b;
    ...
    //direct methods
    ...
    public AdLocusAdapter(AdLocusLayout v) { ... }
    private static WebView a(AdLocusAdapter v) { ... }
    private void a(AdLocusLayout v) { ... }
    ...
    //virtual methods
    public void a() { ... }
    public void b() { ... }
    public void c() { ... }
    ...
}
\end{lstlisting}

In our study, we found an instance of obfuscation where the \emph{WebView} class itself is obfuscated. Various libraries inherit from the \emph{WebView} class and define their own subclass to access \emph{WebView}s functionality. These subclasses are then obfuscated using a code-obfuscator tool. Listing~\ref{listing:obfuscation-adlocus} shows an instance from the library code \emph{Ad-Locus}. Listing~\ref{listing:obfuscation-adlocus} defines a class {\tt AdLocusAdapter} which contains the obfuscated method names such as {\tt a(),b()} and {\tt c()}. To improve precision, static analysis needs to consider these libraries and especially the obfuscation patterns present in libraries.

By adding another layer of complexity to inter-language analysis, obfuscation increases the difficulty for program analysis tools to infer the actual functionality. A precise and sound analysis of these patterns is required for useful analysis results. Obfuscation patterns in Android apps are discussed in detailed in a recent large scale study~\cite{Sascha2018Obfuscation}.

%% file: sections/appendix-js.tex
    \begin{table*}[tb]
        \centering
        \begin{adjustbox}{width=\linewidth}
            \small
    \begin{tabular}{p{10cm}p{1.5cm}p{1.7cm}p{3cm}}
        \toprule
        \textbf{Program Fragment} & \textbf{\%age of Apps} & \textbf{Comments} & \textbf{Randomly Selected Apps}\\
    \midrule
    {\begin{lstlisting}[language=JavaScript,frame=none,aboveskip=0mm,belowskip=0mm,numbers=none]
(function(){var event = document.createEvent('Event');
event.initEvent('fbPlatformDialogMustClose',true,true);
document.dispatchEvent(event);})();
    \end{lstlisting}} & 32.25 & Code from Facebook SDK & com.tivion.usa
    com.com014.lotto
    com.t1gamer.lmcs
    com.ape.games.nlu
    \\
    \hline
    {\begin{lstlisting}[language=JavaScript,frame=none,aboveskip=0mm,belowskip=0mm,numbers=none]
javascript:window.HTMLOUT.
processHTML(document.getElementById('SearchResults').
innerHTML);
    \end{lstlisting}} & 16.65 & Injecting HTML in bridge object \emph{HTMLOUT} & com.www.Fishingmac
    com.tc.veda
    com.alx.roseslwp
    com.mediaapp.mex
    com.calcrate\\
    \hline
    {\begin{lstlisting}[language=JavaScript,frame=none,aboveskip=0mm,belowskip=0mm,numbers=none]
VungleAdjavascript:function actionClicked(m, p) {
    var q = prompt('vungle:' + JSON.stringify({ method: m, params: (p ? p : null) })); 
    if (q && typeof(q) === 'string') { 
        return JSON.parse(q).result; 
    }
}; 
function noTapHighlight() {
    var l = document.getElementsByTagName('*'); 
    for (var i = 0; i < l.length; i++) {
        l[i].style.webkitTapHighlightColor = 'rgba(0,0,0,0)';
    }
};
noTapHighlight();
if (typeof vungleInit == 'function') { vungleInit($webviewConfig$);}
        \end{lstlisting}} & 4.33  & Code from Vungle advertising library & Lucky-Wheel Booboo Guess-The-Flag
        com.ignm.gesu  \\
    \hline
    {\begin{lstlisting}[language=JavaScript,frame=none,aboveskip=0mm,belowskip=0mm,numbers=none]
javascript:window.SynchJS.setValue((function() { 
    try {return JSON.parse( Sponsorpay.MBE. SDKInterface.do_getOffer()).uses_tpn;}
    catch(js_eval_err) {return false;}})()); 
    \end{lstlisting}} & 1.9 & Bridged Communication with Sponsorpay SDK & com.wTieugiano
    SocialNetworkCircus 
    BBCNews-Pashto 
    WoRSA\\ 
    \bottomrule
    \end{tabular}
\end{adjustbox}

    \caption{Most frequently resolved JavaScript code}
    \label{table:top-4-js-code}
    \end{table*}

%% file: sections/evaluation-js-cateogry.tex

In this section we present the insights of our study of \emph{loadUrl} and \emph{evaluateJavascript} on frequently used apps. We collected the most frequently used 196 apps from the Play Store and categorized it into 13 categories. These categories include the most downloaded apps (referred to as Top Apps) and apps that yield high revenues (referred to as Top Gross). Table~\ref{table:js-apps-categories} lists our categories alongside the number of apps studied in each category. 

\begin{table}[tb]
    \centering
    \begin{tabular}{lr}
        \toprule
        \textbf{Category} & \textbf{\#Apps Studied}  \\ \midrule
        Banking & 6  \\
        Business & 10  \\
        Education & 8  \\
        Entertainment & 16  \\
        Health & 10  \\
        Online Payments & 25  \\
        Music & 13  \\
        News & 19  \\
        Shopping & 17  \\
        Social & 11  \\
        Top Grossing & 30  \\
        Top Apps & 22  \\
        Travel & 9  \\ \midrule
        Total & 196  \\ \bottomrule
    \end{tabular}
    \caption{Top apps by categories}  
    \label{table:js-apps-categories}
\end{table}

\begin{figure}[tb]
    \centering
    \includegraphics[scale=0.5]{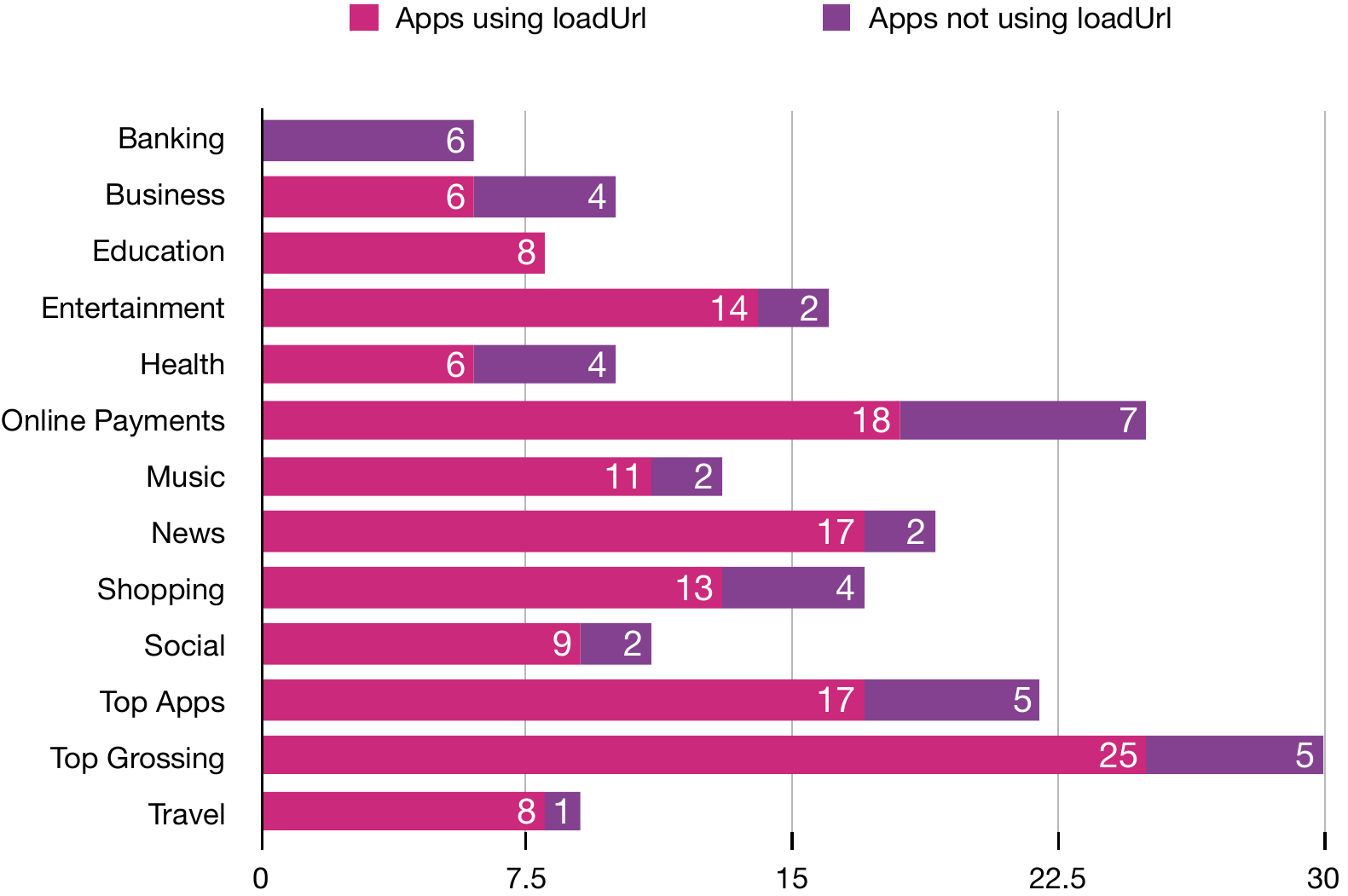}
    \caption{Apps using \emph{loadUrl} in each category}
    \label{fig:app-usage-loadurl-categories}
\end{figure}

\subsection{\JS from \emph{loadURL} in Top 196 Apps}\label{top100-loadURL}
Using \emph{LUDroid}, we prepared a corpus of \JS string and objects' fields (containing a \JS string, potentially a constant) passed to \emph{loadUrl}. We performed a manual analysis of the corpus and answer the questions: \textbf{RQ2.6}, \textbf{RQ3.1 - RQ3.3}. We observed that none of the apps in this category injects third-party scripts remotely over an unsecured protocol (\textbf{RQ3.1}).

\begin{lstlisting}[language=JavaScript,aboveskip=2mm,belowskip=2mm,numbers=none,label={listing:social-fbEvent}, caption={Code from Facebook SDK found in all of the top apps category}, float=tb]
(*\label{listing:social-fbEvent:1}*)(function() { var event = document.createEvent('Event');  
(*\label{listing:social-fbEvent:2}*)    event.initEvent('fbPlatformDialogMustClose',  true,true); 
(*\label{listing:social-fbEvent:3}*)    document.dispatchEvent(event);
})();
\end{lstlisting}
In addition, we also provide insights on the \emph{usage} of \emph{loadUrl} in these apps. Figure~\ref{fig:app-usage-loadurl-categories} lists the percentage of apps using \emph{loadUrl}.
We observed that none of the banking apps we studied use \emph{loadUrl}. 
In this app corpus we founds $50\%$ usage of \emph{loadUrl} in categories other than Banking. These results suggest that a majority of the popular apps, except those related to security-critical banking tasks, leverage \emph{loadUrl}. In what follows, we describe our insights on the usage of \emph{loadUrl} in each of the categories.

In top-grossing and top apps categories, more than $70\%$ of apps use \emph{loadUrl}. On manual inspection, we observed that the use of \emph{loadUrl} in these apps originates from SDKs. Similar to the benign apps in the last section, either the SDKs pass raw \JS or refer to a field variable within the SDK. Approximately $76\%$ of these are mainly related to mobile payments or social networks (\textbf{RQ2.6}). Social networking SDKs account for one-third of the usage among these apps. Listing~\ref{listing:social-fbEvent} lists the most frequent code we observed in this category of apps. Other categories of SDKs prevalent are advertisement (approx. $10\%$), e-commerce (approx. $5.77\%$), and also file sharing SDKs such as \emph{DropBox} (approx. $2\%$). The file-sharing SDK used internally within the \emph{DropBox} app and is used for authentication. These cases highlight the ubiquity of \emph{loadUrl} use among the top downloaded apps and strengthen our results from the benign apps (cf. Section~\ref{subsection:frequent-js}).

In the education apps, such as \emph{EdX} and \emph{Coursera}, we observed that all apps leverage at least one instance of \emph{loadUrl}. We observed a total of 21 instances of its use in 8 apps. Approximately two-third of the apps leverage \emph{loadUrl} provided by the Facebook SDK \textbf{(RQ2.6)} by passing raw \JS or accessing the URL through a field defined in the SDK. Fortunately, none of the SDKs were untrusted \textbf{(RQ2.6)}. This pattern is similar to those observed in benign apps studied in Section~\ref{subsection:frequent-js}. We also observed another a similar pattern of resetting the webpage using \texttt{about:blank} in one app. A minority (approx. $10\%$) of these apps pass the parameter to \emph{loadUrl} through inter-procedural function calls. The usage of \emph{loadUrl} in the entertainment category is similar to those in education. Here, also two-third of the apps leverage \emph{loadUrl}. Additionally we observed the usage of the \emph{vungle} ad library, which was also present in the benign apps.

Next, we consider another category: online payments. We observed that $76\%$ of the apps pass object fields to \emph{loadUrl}, while raw \JS strings account for the remaining $24\%$. All of these object fields are from their respective apps. The raw \JS is a simple facebook login fallback code which is used in case the Facebook app is not installed on the user's device. This is in contrast with our previous observation that object fields passed to \emph{loadUrl} originate from SDKs.

In social apps, we perceive the use of features that modify an app's look-and-feel through Web APIs. Consider the program fragment in Listing~\ref{listing:social-look-and-feel}, where the app uses Web APIs to modify the header section of the app, which is implemented in WebView. Line~\ref{listing:social-look-and-feel:1} selects the element named \emph{MTopBlueBarHeader}, and  line~\ref{listing:social-look-and-feel:2}  removes all styling. In this category we found two unique type-2 source code clones, and no instances of passing objects' fields to \emph{loadUrl}. 
\begin{lstlisting}[language=JavaScript,aboveskip=2mm,belowskip=2mm, label={listing:social-look-and-feel}, caption={Modifying an app's look-and-feel}, float=tb]
(*\label{listing:social-look-and-feel:0}*)(function() { 
(*\label{listing:social-look-and-feel:1}*)  var topbar = document. querySelector('#header[data-sigil = "MTopBlueBarHeader"]'); 
(*\label{listing:social-look-and-feel:2}*)  topbar.setAttribute('style', 'display:none');
(*\label{listing:social-look-and-feel:3}*)})()
\end{lstlisting}

In other categories, such as shopping and travel, we observed a single type-2 clone of raw \JS shown in Listing~\ref{listing:social-fbEvent}. The same code snippet is most frequent among benign apps as well (Section~\ref{subsection:frequent-js}). Line~\ref{listing:social-fbEvent:2} initializes a DOM event handler that listens to the event fbPlatformDialogMustClose, and line~\ref{listing:social-fbEvent:3} dispatches the corresponding event.

\subsection{Use of \eJS in frequently used apps \label{section:top-100-js}}
\input{sections/evaluate-ejs.tex}

%% file: sections/evaluate-ejs.tex
\begin{table*}[tb]
    \centering
        \begin{tabular}{lrl}
        \toprule
        \textbf{Program Fragment} & \textbf{\%age of Apps} &  \textbf{Comments}\\
        \midrule
        {\begin{lstlisting}[language=JavaScript,frame=none,aboveskip=0mm,belowskip=0mm,numbers=none,escapeinside={@}{@}]
document.loginForm.j_password.value = "@\dots@";
        \end{lstlisting}} & 9 & \multirow{6}{*}{Injects Sensitive Data} \\   \cline{1-2}
        {\begin{lstlisting}[language=JavaScript,frame=none,aboveskip=0mm,belowskip=0mm,numbers=none,escapeinside={@}{@}]
document.getElementById('access-pin').value = "@\dots@";
        \end{lstlisting}} & 9 & \\ \cline{1-2}
        {\begin{lstlisting}[language=JavaScript,,frame=none,aboveskip=0mm,belowskip=0mm,numbers=none,escapeinside={@}{@}]
document.getElementById('access-code').value = "@\dots@";
        \end{lstlisting}} & 9 & \\ \cline{1-2}
        {\begin{lstlisting}[language=JavaScript,frame=none,aboveskip=0mm,belowskip=0mm,numbers=none,escapeinside={@}{@}]
document.logon1.PASSWORD1.value = "@\dots@";
                    \end{lstlisting}} & 9  &  \\ \cline{1-2}
        {\begin{lstlisting}[language=JavaScript,frame=none,aboveskip=0mm,belowskip=0mm,numbers=none,escapeinside={@}{@}]
document.form1.selBankID.value = "@\dots@";
        \end{lstlisting}} & 5  & \\ \cline{1-2}
        {\begin{lstlisting}[language=JavaScript,frame=none,aboveskip=0mm,belowskip=0mm,numbers=none,escapeinside={@}{@}]
document.form1.consumerEmail.value = "@\dots@";
        \end{lstlisting}} & 2  &  \\ \bottomrule
        \end{tabular}
    \caption{JavaScript code passed to \eJS. Sensitive values are masked by \texttt{\dots}  to protect the confidentiality of the information in these apps.}
    \label{table:program-js-sensitive}
\end{table*}
We also analyzed the most frequently used apps from the Google PlayStore to study the \JS code passed to the method \eJS. In particular, we found differences between the usage of \eJS and \emph{loadUrl} in these apps. 
In Section~\ref{section:top-100-js}, we identified some usage patterns of \emph{loadUrl}, many of which similar to the benign apps in Section~\ref{subsection:frequent-js}. However, in the top 196 apps, our study found a significant difference between the usage of \eJS and \emph{loadUrl}. We found that 26 of these 196 apps use at least one instance of \eJS.
These usages contain 158 instances of \JS code fragments, forming 16 type-2 code clones. In what follows, we elaborate two cases of frequent \JS code in these apps. We also observe an unusual case which -- to our surprise -- violates basic security practices.

\subsubsection{Case Study --- Controlling Page Navigation} 
In our study, we found apps using \JS to control page navigation on Android, which is similar to using a regular web page. Consider the code snippet from the payment apps \emph{JioMoney} and \emph{LiquidPay}, where the developers use handlers for keyboard events in \JS: \lstinline[language=JavaScript]|handleBackKey()|, \lstinline[language=JavaScript]|LoginSubmitClick()|, and \lstinline|document.getElementById('loginIdxSubmitBtn').click()|. The first one registers a custom handler for handling the back key on Android devices, while the second one is for logging in via a submit button, and the third one automatically initiates a click event. This case study highlights sophisticated use of \eJS corresponding to the sophisticated use in \emph{loadUrl} (cf. Section~\ref{subsection:frequent-js} and Section~\ref{section:top-100-js}).

\subsubsection{Case Study --- LiquidPay} Table~\ref{table:program-js-sensitive} displays six code snippets we identified using \emph{LuDroid}. The payment app \emph{LiquidPay} uses sensitive information in plain-text in five of its Activities. All of these five Activities use sensitive data such as a password or Bank-ID or perform a sensitive operation such as User Login. The developers include sensitive values in plain-text such as \lstinline{document.getElementById('access-code').value = "\dots"} or \lstinline{document.form1.selBankID.value = "\dots"} in \emph{WebView}~(cf. Table~\ref{table:program-js-sensitive}). We have masked the sensitive information by \texttt{\dots} to protect the confidentiality of the information of this app. This case study exemplifies a misuse of \emph{WebView}.

To the best of our knowledge, we did not find any relevant work on static analysis of \eJS. Our study on the most frequently used apps dataset reveals that developers have found sophisticated use cases for \eJS. In principle \eJS provides an alternative to \emph{loadUrl}. Although \eJS is meant to only evaluate the passed \JS expression, it is also possible to mimic the functionality of \emph{loadUrl} by manipulating DOM properties. Again, this makes the existing analyses~\cite{lee2016hybridroid,Ryu2019} unsound. A more precise and sound analysis has to also consider the data-flow between Android and \eJS.

%% file: sections/evaluation-malware.tex
We manually analyzed 1000 malware samples and studied the \JS passed to \emph{loadUrl} and \eJS. To our surprise, we did not find any instances of \eJS use in malware. Of all malware samples we studied, 121 samples pass \emph{\JS} code to \emph{loadUrl}, while 451 samples use \emph{loadUrl} with an object's field variable. The 121 instances of raw \JS constitute eight distinct code clones (type~2) . We further performed a manual analysis of these eight unique \JS codes and found two of these \JS codes to be similar (type~2 clone) to those in the benign apps. A manual analysis of the field variables again revealed that these originate from SDKs, similar to those we found in benign apps (cf. Table~\ref{table:sdk-usage}).

\subsection{Case Study --- Injecting external objects \label{subsubsection:injecting-malware}}
We found a usage pattern in 88 of the 121 distinct malware codes (approx $72.73\%$), where the malware injects results from the WebView into a shared Android object. Listing~\ref{listing:android-malware-code-injection} shows one such instance where the DOM element named \emph{SearchResults}, is passed to the method \emph{processHTML}, and the resulting HTML is accessed through the bridge object \emph{HTMLOUT}. This pattern is similar to injecting external results in Android in Section~\ref{subsection:frequent-js}. However, in this case, it is more dangerous. By manually searching for the package names, we find approximately $63\%$ of the malware are from repackaged apps. It points out that these apps either have a malicious activity or \emph{WebView}. Having a malicious \emph{WebView} is more dangerous as it potentially exposes it to the Web.

Further, the activities (or WebView) in these apps run with the same privileges as a regular app. For the other $37\%$ apps, we could not find any information on the Google Play Store. Therefore, these apps are taken down by the user, or otherwise, these apps performed some suspicious operation. However, with the current data that we have, it is not possible to give further comments.

\begin{lstlisting}[language={JavaScript},aboveskip=2mm,belowskip=2mm, label={listing:android-malware-code-injection}, caption={Malware injecting external results in Android object}, float=tb]
window.HTMLOUT.processHTML(  document.getElementById('SearchResults').  innerHTML);
\end{lstlisting}

\subsection{Case Study --- Clickjacking \label{subsubsection:clickjacking-malware}}
As another case study we identified is a clickjacking attack. Listing~\ref{listing:android-malware-clickjacking} shows a code snippet found in five out of 121 malware (approx. $4.13\%$ ). Similar to the case study in Section~\ref{subsection:frequent-js}, this code snippet is also susceptible to a clickjacking attack. The function \emph{noTapHighlight} in Listing~\ref{listing:android-malware-clickjacking} masks user-clicks by removing the feedback given to the users (by removing the highlighted color), thus, making them unaware of the accidental user-interactions with the device. Line~\ref{listing:android-malware-clickjacking:1} of the function \emph{actionClicked} serializes the arguments passed to the method. Line~\ref{listing:android-malware-clickjacking:2} returns the value of the field \emph{result} after a deserialization of the previous statement. We discover the source code in malware, and therefore, the intent of the clickjacking attack could potentially be malicious. However, owing to the limitations of \emph{LuDroid}, we could not ascertain the exact sequence of user-interactions that invokes these functions.

\begin{lstlisting}[language=JavaScript,aboveskip=2mm,belowskip=2mm, label={listing:android-malware-clickjacking}, caption={Malware susceptible to clickjacking},float=tb]
function actionClicked(t, u) { 
(*\label{listing:android-malware-clickjacking:1}*)    var r = prompt('showToast' + JSON.stringify({ method: 'showToast', params: (u ? [t, u] : [t]) })); 
(*\label{listing:android-malware-clickjacking:2}*)    if (r && typeof r === 'string') { return JSON.parse(r).result; } 
};
function noTapHighlight() { 
    var l = document.getElementsByTagName('*'); 
    for (var i = 0; i < l.length; i++) { 
        l[i].style.webkitTapHighlightColor = 'rgba(0,0,0,0)'; 
}}; 
noTapHighlight();
\end{lstlisting}

The source code mentioned here is a type-2 clone of the source code in Listing~\ref{listing:leaking-information}, which was also observed in the benign apps in Section~\ref{sec:vungle}. However, it makes the scenario more dangerous than those in benign apps as the same technique is applied to apps with malicious intent. 
This case study reveals the severity of vulnerable patterns (Section~\ref{subsection:frequent-js}) if exploited by malicious apps. We attempted to map the respective source codes between benign apps and malware. Unfortunately, a similarity could not be established owing to the limitations of decompilation. Decompilation mangled the variables and class names in many cases, so building a one-to-one mapping between malware and benign apps is not feasible.

\subsection{Case Study --- Analytics SDK in malware}
We identify a case study where, to our surprise, we found malware using the analytics SDK \emph{Flurry}. Consider Listing~\ref{listing:android-malware-flurry} where the function defines an adapter to listen to the Flurry analytics services. The body of the anonymous function implements the functionality possibly required for handling of call requests stored in the call queue. We found this pattern in 8 of 121 apps (approx. $6.61\%$) of apps. 
\begin{lstlisting}[language=JavaScript,aboveskip=2mm,belowskip=2mm, label={listing:android-malware-flurry}, caption={Malware using SDK},float=tb]
(function() {
    var flurryadapter = window.flurryadapter = {};       
    flurryadapter.flurryCallQueue = [ ];       
    flurryadapter.flurryCallInProgress = false;       
    flurryadapter.callComplete = function(cmd) {          
        if ( this.flurryCallQueue.length == 0 ) {             
            this.flurryCallInProgress = false;             
            return;          
        }
        var adapterCall = this.flurryCallQueue.pop();           
        this.executeNativeCall(adapterCall);          
        return "OK";       
    };       
    //Remaining code from Flurry SDK
})();
\end{lstlisting}

\subsection{Case Study --- SDK usage in malware}
We also identified a case where malware uses SDKs, especially those SDKs concerned with advertising. The SDK usage pattern is prevalent in $48.1\%$ of the studied malware samples. Out of these, we found $78.2\%$ (approx.\@ $37\%$ of all malware) originating from SDKs meant for advertising. For example, consider the use of the advertising library \emph{TapJoy} in malware \texttt{com.nuttyapps.sally.makeup.salon} and \texttt{skloo.mobile.pud}. Note that the former APK is probably derived from a game called ``Sally's Makeup Salon'', the latter an ``Ultimate Drinking'' app. Malware is often piggybacked to popular benign software and distributed via non-regulated app stores as a trojan horse in order to be installed voluntarily by naive users.
In related work, Lee and Ryu have shown that the TapJoy library performs various sensitive operations. These include launching new activities, retrieving sensor information, and fetching location information and app information~\cite{adLib}. The last three operations potentially harm user privacy. However, this case is more dangerous because of the malicious intent of the built-in malware. Launching new activities through \JS leads to a \emph{App to Web Injection attack}~\cite{hassanshahi2015DESORICS},~\cite{adLib} where the app (in this case, malware) initiates new malicious activities through injected \JS.

A closer investigation of which functionality stems from the regular application and in which form the attached malware is beyond the scope of this paper. However, we can see that malware leverages hybrid app activities in order to satisfy its malicious intents and that analyses that aim at analyzing the hybrid communication are bound to provide a precise model of the communication patterns detected in this study. Therefore our work may also serve as a ``testbed'' for such analyses.

\subsection{Comparison of JavaScript Usage Across Datasets}
 
In Section~\ref{methodology-js}, we began with three research questions concerning the use of third-party script injection over unsecured protocols \textbf{(RQ3.1)}, non-trivial information flows  from \JS to Android, and third-party libraries using obfuscation in their \JS code. Besides, we also discovered the vulnerable use case of leaking private information through clickjacking. In this section, we summarize our observations on the \JS strings passed to \emph{loadUrl} for the three categories of apps: Benign, Frequently used, and Malware.

\begin{table}[tb]
    \centering
    \begin{tabular}{p{0.3\columnwidth}ccc}
        \toprule
        \textbf{Category} & \textbf{Benign} & \parbox{0.2\columnwidth}{\textbf{Frequently Used}} & \textbf{Malware} \\ \midrule
        Third-party script injection & \cmark & \xmark & \xmark \\ \hline
        Non-Trivial Information Flow  & \cmark & \xmark & \cmark \\ \hline
        Obfuscation by Third-Party  & \cmark & \xmark & \xmark \\ \hline
        Privacy leak & \cmark & \cmark & \cmark \\
        \bottomrule
    \end{tabular}
    \caption{Summary of  the observed \JS behavior in studied apps}
    \label{table:summary-observed}
\end{table}

Table~\ref{table:summary-observed} presents the behavior observed across the datasets of apps. We observed that (potential) privacy leak is a prevalent behavior of the \JS code passed to \emph{loadUrl} in all of the app categories. Non-trivial information flows from \JS to Android is the next typical behavior, observed only in the benign apps and malware dataset. Interestingly, the frequently used apps do not show this behavior as they, primarily, use it through field objects (mostly constant strings) of the SDKs. Obfuscated third-parties were observed only in benign apps. 
Furthermore, frequently used apps and malware were free from third-party script injection attacks. It shows that at least the frequently used apps adhere to basic security practices.

%% file: sections/discussion.tex
LUDroid was able to derive a plenitude of novel statistics and case studies examining information flows, URL statistics and statistics on JavaScript code. At this point LUDroid is not a stand-alone analysis tool but merely supports manual inspection and calculation of statistical data. The goal of this work is to present interesting insights on \emph{how the bridge between Android and JavaScript is used in the wild}, in order to capacitate the design of automatic program analyses that take both sides of the hybrid app into account.
Therefore, it cannot be the aim of this work to analyze the plentitude of \JS code loaded through HTML files, which requires automatic analysis due to the sheer size of the code base.

Obviously the data we gathered depends on the corpus of apps in the study, and there is a risk that the investigated apps are not representative. However, due to the fact the we randomly chose 7500 apps from a database crawled between 2015 and 2019 should guarantee that we are not biased to any particular app format. To even out the randomness we are adding the most popular apps and malicious apps.

Other threats to the validity of our study stem from the limitation of our approach, which we discuss in the sequel.

\subsection{Limitations of LUDroid}
\emph{LuDroid} leverages static analysis techniques and therefore inherits some of the typical challenges and limitations. As a tool that supports manual investigation of bridge communication we are currently not interested to handle challenging topics like native code, reflection, dynamic control flow, obfuscation, and the fact that strings like the URLs passed to \emph{loadURL} can be constructed at runtime. All of these obstacles have been investigated in separate lines of research~\cite{DBLP:conf/esorics/GrossT018,iifa,android-native,android-reflection-lux,heapsYannnis}
, and we consider them orthogonal to the insights we are aiming at in this study. Note, however, that \emph{loadURL} is also a dynamic language feature that, like reflection, may execute code constructed at runtime based on a string parameter. Insights gained in studies that target dynamic code execution (e.g.~for JavaScript~\cite{eval}) are also relevant to understand the semantics of \emph{loadURL} and \eJS. A fully implemented static analysis that automatically derives information flows across the language barrier will eventually want to at least approximate these challenging features.

\paragraph{Native code} In Android applications, it is possible to include native code (e.g., compiled C/C++ code) via the Java Native Interface (JNI). Including assembler semantics into the analysis increases its complexity significantly. Fortunately, we have observed native code rarely during our experiments, so we do not expect a lot of loss in terms of the goals of this study.

\paragraph{Reflection, dynamic control flow and obfuscation} Reflection in Java is a means to access code features at runtime. With reflection, it is possible to make calls and access fields at runtime, dynamically depending on strings and other values derived at runtime. It is, therefore, possible to make the control and data flow entirely dependent on runtime values, therefore eluding static analysis. At present, \emph{LUDroid} does not support analysis for reflective features. Given this limitation, LUDroid will lack insights from apps that intentionally use reflection to prevent static analysis (e.g. trough obfuscation). Again, we have not encountered high usage of reflection in any of the inspected benign and maliciouas apps.

\paragraph{Limited String analysis} LUDroid requires string analysis to resolve arguments passed to the \emph{loadURL} and \eJS methods. For cases where the string argument is manipulated before being passed, we implemented domain knowledge of the Java \emph{String} class as well as support for the \emph{StringBuilder} class to be able to statically derive the results of straightforward string manipulation. However, LUDroid does not support advanced string manipulations such as array-based string manipulations at the current point of time. More advanced resolution techniques are envisioned for future analyses.

\paragraph{InterProcedural slicing} At the time of this writing LUDroid only supports intra-procedural slicing. For the fast dissemination of typical use cases of bridge communication this design decision was found to be sufficient, as we were not planning to be able to present a complete picture of communication patterns. Future analyses will have to consider more complex communication scenarios to be able to evaluate the security and/or privacy properties of an app under investigation.

%% file: sections/related-work.tex
Rizzo et al.~\cite{rizzo2017babelview} proposed BabelView, which models JavaScript as a blackbox. They leverage static taint analysis to detect unwanted information flows and  five different vulnerability types.  Zhang et al.~\cite{zhang2018empirical} performed a large scale study of the WebView APIs to classify them into four categories of web resource manipulation. Hidhaya et al.~\cite{hidhaya2015supplementary} described the "supplementary event-listener injection attack" in Android WebViews. They further proposed a tool for automated detection of this vulnerability and a mitigation. Li et al.~\cite{li2017unleashing} discovered a new type of WebView-based attack that they call Cross-App WebView Infection (XAWI).  Mandal et al.~\cite{mandal2018vulnerability} proposed a static analysis tool to detect various vulnerabilities in Android Infotainment applications. Their approach is based on Julia, a static abstract interpretation analysis tool. Fratantonio et al.~\cite{Fratantonio2016LogicBomb} proposed a static analysis tool to detect malicious application logic in Android apps. Their approach is based on various known static analysis techniques such as symbolic execution and inter-procedural control-dependency analysis. In contrast to these approaches, our work is not limited to specific vulnerabilities, but provides useful insights by inspecting both Android and JavaScript code.

Lee et al.~\cite{lee2016hybridroid} proposed HybriDroid, an information flow analysis tool based on WALA. They discussed the semantics of WebView communication including type conversion semantics between Java and JavaScript. In contrast to our work, HybriDroid does not provide a comprehensive analysis of hybridization APIs. Their approach is restricted to the fundamental taint analysis of the Information flow from Android to JavaScript and misses other valuable insights. In contrast, we present the full picture of non-trivial data and control flows that occur in Android Web-Hybridization. 
Besides, we provide experience of the usage of URLs and exemplarily exploit the insecure usage of URLs.  Our insights aim to improve the unsoundness in the existing analyses and thus benefit further research.

Neugschwandtner et al.~\cite{Neugschwandtner2013exploitation} proposed two attack scenarios based on when the client or 
server is compromised. Their approximation is quite coarse in case of privacy leakage where a trusted 
channel could leak more than the required information. 
Mutchler et al.~\cite{Mutchler2015large-scale-study} conducted a large-scale study of apps using WebView aiming at the security vulnerabilities present in these apps. However, this study focuses only on particular types of vulnerabilities and they did not consider the misuse of JavaScript in \texttt{loadURL}. 
Yang et al.~\cite{yang2018study} examined so called ``Origin Stripping Vulnerabilities'' caused by wrongly using the \emph{loadURL} method. 
 
Bae~\cite{Ryu2019} formalized the semantics of the android interoperations between Java and JavaScript. Their approach proposed a type-system based error detection for \verb|MethodNotFound| errors. However, their approach does not consider the information flow from JavaScript to Android Java.
In addition to a large scale study, Kim et al.~\cite{Kim2012Scandal} leveraged abstract interpretation to design a static 
analysis that finds privacy leaks in android applications. Targeting excess authorization and file-based cross site scripting attacks, Chin et al.~\cite{Chin2013vunerabilities} proposed Bifocals, a tool to detect these vulnerabilities. However, these  analyses focused on one particular part of the problem. Our study is targeted at analyzing all programming patterns which may potentially lead to vulnerabilities. 

In general the analysis of unencrypted communication in Android apps is a well-explored topic~\cite{pokharel2017can, he2014security, fahl2012eve}. For example, Pokharel et al.~\cite{pokharel2017can} demonstrated eavesdropping attacks on VoIP apps. However, to the best of our knowledge no previous work has analyzed the security consequences of unencrypted communication caused by \emph{loadURL}.

%% file: sections/conclusion.tex
In this work, we present a large-scale analysis of \emph{loadURL} and \eJS usages in real world applications. The statistical results include numerous features, such as information flow data, URL statistics and JavaScript code features on a set of 7,500 randomly selected applications from the Google Playstore, the most popular apps and 1000 malware samples. We implemented our semi-automatic analysis approach in a tool called LUDroid that computes the data by using slicing techniques. 
LUDroid discovered many instances of vulnerabilities, e.g. concerning the usage of unprotected protocols in URLs. To demonstrate the validity of these vulnerabilities we exemplarily showcased the exploitation of one of them. Investigating the most popular apps and malware samples displayed differing characteristics than those found in general benign apps. The insights gained in this study provide valuable input for designing program analyses that are to analyze hybrid Android apps.